\documentclass[aps,showpacs,preprint,preprintnumbers,amsmath,amssymb,nofootinbib]{revtex4-1}
\usepackage{graphicx}
\usepackage{dcolumn}
\usepackage{bm}
\usepackage{siunitx}
\usepackage{slashed} 
\usepackage{subfigure}
\newcommand{\bea}{\begin{eqnarray}}
\newcommand{\eea}{\end{eqnarray}}

\def\missE{\slashed E} 

\begin{document}

\title{Probe Higgs boson pair production via the  $3 \ell 2 j$ +  ${\slashed E}$ mode}
\author{Qiang Li$^{1,2}$, Zhao Li$^{3}$, Qi-Shu Yan$^{2,4,5}$, Xiaoran Zhao$^{3,4} \footnote{Correspondence Author:zhaoxiaoran13@mails.ucas.ac.cn}$
\\$^{1}$ Department of Physics and State Key Laboratory of Nuclear Physics and Technology, Peking University, Beijing, 100871, China
\\$^{2}$ CAS Center for Excellence in Particle Physics, Beijing 100049, China
\\$^{3}$  Institute of High Energy Physics, Chinese Academy of Sciences Beijing 100039, P.R. China
\\$^{4}$ School of Physics Sciences, University of Chinese Academy of Sciences, Beijing 100039, P.R. China
\\$^{5}$ Center for High-Energy Physics, Peking University, Beijing, 100871, P.R. China
}
\begin{abstract}

We perform a detailed hadron-level study on the sensitivity of Higgs boson pair production via the $WW^{*}WW^{*}$ channel with the final state $ 3 \ell  2j + {\slashed E}$ at the LHC with the collision energy $\sqrt{S} = 14$ TeV and a future 100 TeV collider. To avoid the huge background from $pp \to Z W + \textrm{jets}$ processes, we confine to consider the four lepton patterns: $e^\pm e^\pm \mu^\mp $ and $\mu^\pm \mu^\pm e^\mp$. We propose a partial reconstruction method to determine the most reliable combination. After that, we examine a few crucial observables which can discriminate efficiently signal and background events, especially we notice that the observable $m_{\rm T2}$ is very efficient. For the LHC 14 TeV collisions, with an accumulated 3000 fb$^{-1}$ dataset, we find that the sensitivity of this mode can reach up to 1.5 $\sigma$ for the Standard Model and the triple coupling of Higgs boson $\lambda_3$ in the simplest effective theory can be constrained into the range [-1, 8] at $95\%$ confidence level; at a 100 TeV collider with the integrated luminosity 3000 fb$^{-1}$, the sensitivity can reach up to 13 $\sigma$ for the Standard Model and we find that all values of $\lambda_3$ in the effective theory can be covered up to 3$\sigma$ even without optimising signals. To precisely measure the triple coupling of Higgs boson $\lambda_3=1$ of the Standard Model at a 100 TeV collider, by using the invariant mass of three leptons which is robust to against the contamination of underlying events and pileup effects and by performing a $\chi^2$ analysis, we find that it can be determined into a range [0.8, 1.5] at $95\%$ confidence level.
\end{abstract}


\maketitle

\section{Introduction}

The last building block of the Standard Model (SM), Higgs boson, has been discovered by ATLAS and CMS Collaborations \cite{Aad:2012tfa,Chatrchyan:2012ufa}. The interaction of Higgs boson with the fermions of the SM and its self couplings are new types of interactions which are different from those described by the gauge symmetries in the SM. To ascertain the nature of Higgs boson, it is important to precisely measure the Yukawa type interactions which can be determined by measuring the Higgs decay into fermion pairs from single Higgs production at future LHC runs and Higgs factories\cite{Baer:2013cma,Dawson:2013bba}. While the analysis on the Higgs self couplings via Higgs pair and multi-Higgs boson production is achievable at high luminosity LHC runs and future pp collider, say a 100 TeV collider \cite{Brock:2014tja}. 

The determination of the Higgs potential is of important significance, since the potential is directly related to the structure of vacuum, the electroweak phase transition and electroweak baryongensis, and the fate of our universe as well. It is useful to address the issue whether the Higgs boson is elementary or composite. It is also crucial to probe new physics, which is believed to exist somewhere and somehow since there are fundamental issues which cannot be solved by the SM itself, e.g. the matter-antimatter asymmetry in our universe, the quadratic divergence of the Higgs mass term, and the mystery of dark matter, etc.

The SM predicts trilinear and quartic self couplings in the Higgs potential at tree level. Both trilinear and quartic Higgs self couplings are related to the Higgs boson mass by $m_H^2 = \frac{1}{2} \lambda_{\textrm{SM}} \, v^2$, where trilinear and quartic couplings are proportional to $\lambda_{\textrm{SM}}$, which is the dimensionless coupling of Higgs potential before electroweak symmetry breaking. In the language of an effective field theory, the trilinear coupling term can be simply expressed as 
\bea
L = \frac{\lambda_3}{6} \,\lambda_{\textrm{SM}}^{tr} \,\,v\, H^3 = \lambda_3 \,\, \frac{\lambda_{\textrm{SM}}}{4} \,v \,H^3\,,
\eea
where $\lambda_3=1$ corresponds to the SM case and there is relation $\lambda_{\textrm{SM}}^{tr} = 3/2 \lambda_{\textrm{SM}}$ in this parametrisation. It is well known that to determine the quartic coupling of the SM might be challenging at the LHC due to the small production rate of three Higgs boson final state, but to detect the trilinear coupling via Higgs pair production is expected to be within the reach of the LHC. The measure of the trilinear coupling up to the precision $10\%$ at a future 100 TeV colliders is feasible \cite{Yao:2013ika}, which can further pinpoint and discover new physics.

The importance of Higgs pair production has attracted attentions long time ago. Theoretical investigations on the Higgs pair production in the SM began with the pioneering works \cite{Eboli:1987dy,Keung:1987nw,Dicus:1987ez}, where the gluon-gluon fusion \cite{Eboli:1987dy} and the vector boson fusion \cite{Keung:1987nw,Dicus:1987ez} processes had been considered. It has been found that at hadron colliders the gluon-gluon fusion production is almost one order of magnitude larger than the weak boson fusion process. There are lots of effort to improve theoretical prediction on the Higgs pair production. For example, the NLO and NNLO QCD corrections to gluon-gluon fusion had been considered in \cite{Plehn:1996wb,Dawson:1998py} and recently in \cite{deFlorian:2013jea} by using the large top mass approximation and normalizing the partonic cross section using the exact LO result. The finite top quark mass effects have been analysed at NLO in \cite{Grigo:2013rya} via expansion by top quark mass. Recently,  NNLO QCD corrections to the VBF Higgs pair production has been done by the USTC group\cite{Liu-Sheng:2014gxa}. 

Besides the detecting of the Higgs self-couplings of the SM, multi-Higgs production at various colliders are of great importance to probe new physics, as explored in reference \cite{Asakawa:2010xj}. At hadron colliders, Higgs pairs can be enhanced by other heavier scalar resonances \cite{Kribs:2012kz,No:2013wsa,Liu:2013woa,Heng:2013cya,Cao:2014kya}. By measuring the signal of Higgs pair production, we can extract the triple Higgs coupling and then depict the shape of Higgs potential so as to distinguish various electroweak symmetry breaking models. For example, the composite models predict a vanishing or small triple couplings \cite{Doff:2009na} and a model with effective potential $V= \lambda (H^+ H)^2 + (H^+ H)^3/\Lambda^2$ predicts a triple Higgs coupling $7/3$ times that of the SM. The measurement of the cross section of Higgs pair production is also important to distinguish models where Higgs is assumed to be elementary, like in the supersymmetric model where superparticles can enhance the production rate \cite{Belyaev:1999mx,Belyaev:1999vz,Cao:2013si} and like in the two-Higgs doublet model the extra scalars can enhance the production rate \cite{Moretti:2004wa} . While the Higgs-Gravity model \cite{Xianyu:2013rya,Ren:2014sya} predicts a coupling dependent of external momenta. These specific models can be more generally formulated and conveniently explored in the framework of the effective Lagrangian up to $O(p^6)$\cite{Giudice:2007fh}, as demonstrated in a recent study in Reference \cite{Goertz:2014qta}. 

A comprehensive study on various productions at the generator level has been recently investigated in \cite{Frederix:2014hta} by using the automatic matrix element generator Madgraph5. According to the study of \cite{Frederix:2014hta}, in the SM the leading contribution to Higgs pair production at the LHC and a future 100 TeV collider is via gluon-gluon fusion. The subleading production mechanism is via weak vector boson fusion processes \cite{Baglio:2012np,Liu-Sheng:2014gxa}. The $t{\bar t}$ associated production can become comparable with the weak vector boson fusion production when the collision energy is around 100 TeV  \cite{Frederix:2014hta}. The effects of top quark mass in double and triple Higgs production at hadron colliders have been studied in \cite{Maltoni:2014eza}.  The kinematics of the di-Higgs bosons decay to $b{\bar b}\gamma\gamma$ have been analyzed in \cite{Slawinska:2014vpa} and their effects to the measurement of non-standard values of $\lambda_{3}$ have been explored. Interested readers can refer \cite{Shao:2013bz,Li:2014yta} for theoretical progress in the fixed order QCD calculation for single Higgs and Higgs pair productions, and top quark pair production as well.

Except the theoretical efforts on the Higgs pair production, there are lots of efforts to improve theoretical predictions of the Higgs boson decay. For a Higgs boson with mass around 125-126 GeV, its main decay final state is $b\bar{b}$, and the state-of-art research on its partial width is up to $O(\alpha_S^4)$ in \cite{Baikov:2005rw}. Higgs decaying into other fermion pair have also been investigated up to two loop level. The $H\to gg$ decay channel is up to N$^3$LO QCD in \cite{Baikov:2006ch} in the large top quark mass limit, and the top quark mass effects are analyzed in \cite{Schreck:2007um}. The partial width of $H\to \gamma\gamma$ channel is known up to NLO EW and NNLO QCD\cite{Passarino:2007fp,Maierhofer:2012vv}. The decay channel $H\to Z\gamma$ is known up to NLO QCD in \cite{Spira:1991tj}. For the decay channel $H\to WW^{*},ZZ^{*}\to 4f$, $O(\alpha_S)$ and $O(\alpha)$ corrections have been studied in \cite{Bredenstein:2006rh,Bredenstein:2006ha,Bredenstein:2007ec}. Interested readers can refer \cite{Djouadi:2005gi,Djouadi:2005gj,Dittmaier:2011ti,Dittmaier:2012vm,Heinemeyer:2013tqa} for more information on the current status of our understanding to Higgs boson.

Recently the signals of Higgs boson pair production at the LHC have been further studied via a few decay channels. A recent theoretical review can be found in \cite{Baglio:2012np,Baglio:2014xja}.  For example, the study of $H H\to b\bar{b}\gamma\gamma$ channel can be found in Ref. \cite{oai:arXiv.org:hep-ph/0310056} with a significance of about 1.5 $\sigma$ for a integrated luminosity $\SI{600}{fb^{-1}}$ at LHC with 14 TeV collision energy is assumed. A recent search by the CMS collaborations can be found in Ref. \cite{CMS:2014ipa}. The authors of the Ref. \cite{Baglio:2012np} updated this study and provided a significance of about 6.46 $\sigma$ for the integrated luminosity $\SI{3000}{fb^{-1}}$ at the 14 TeV. The study for the $H H \to b\bar{b}\tau^{+}\tau^{-}$ channel can be found in Ref.\cite{oai:arXiv.org:hep-ph/0304015,oai:arXiv.org:1206.5001,Baglio:2012np}, where the authors of the  Ref.\cite{Baglio:2012np} provided a significance of about 9.36 $\sigma$ for $\SI{3000}{fb^{-1}}$ LHC. The channel $H H \to b\bar{b}W^{+}W^{-}\to b\bar{b}\ell \nu_\ell jj$ has been studied in Ref.\cite{oai:arXiv.org:1209.1489}, where a significance of 3.1 $\sigma$ for $\SI{600}{fb^{-1}}$ LHC have been obtained. The mode $H H \to b\bar{b}W^{+}W^{-}\to b\bar{b}\ell \nu_\ell \ell' \nu_{\ell'}$ has been studied in Ref.\cite{Baglio:2012np} and a significance of about 1.53 $\sigma$ for $\SI{3000}{fb^{-1}}$ the LHC has been achieved. A recent updated study on 4 b jet final state can be found in Ref. \cite{deLima:2014dta} and a search for new physics by the CMS collaborations can be found in Ref. \cite{Khachatryan:2015yea}. The probe of the VBF Higgs pair production can be found in Ref. \cite{Dolan:2013rja}.


The third largest decay fraction channel for Higgs pair is $WW^{*}WW^{*}$ channel. The subsequent decay mode $8j$ and $l 6j+\missE$ will be too hard to be found due to huge QCD multi-jets and W+multi-jets background. The decay mode $l^{\pm} l^{\mp}4j +\missE$ will be also too hard to be found due to the huge Z($\gamma^{*}$)+multi-jets and $W^{+}W^{-}$+multi-jets background. The decay mode $4l +$ missing energy will has a tiny production rate. So only two subsequent decay channels are reachable: two same-signed lepton mode $l^{\pm} l^{\pm}  4j + \missE$ and three leptons mode $3 \ell 2j + \missE$. 

These channels had been taken into account in Ref. \cite{oai:arXiv.org:hep-ph/0206024,Baur:2002qd} with the assumption that the Higgs mass is in the range $\SI{140}{GeV} < m_H < \SI{200}{GeV}$ and the $WW$ is the main decay channel for Higgs and both W bosons are on shell, where, at parton level, important acceptance cuts and some simple kinematic variables, especially the invariant mass of all final states which was crucial to suppress the background events of $t \bar{\textrm{t}} + \textrm{jets}$ and multi-top processes, were carefully studied. Considering that the measured Higgs boson mass is 125 GeV or so and the branching fraction of Higgs boson decay to $W W^{*}$ is considerably smaller than that assumption in Ref. \cite{oai:arXiv.org:hep-ph/0206024,Baur:2002qd}, the production rate of signal in this final state is almost one order of magnitude smaller and the discovery of the signal in this mode is very challenging. Furthermore, not all the W bosons from the decay of Higgs boson can be on-shell which makes the signal hard to be distinguished. Therefore it is necessary and quite nontrivial to revisit and perform a more detailed analysis by taking all these facts into account. In this work, we propose a partial reconstruction procedure of Higgs pair in the final states and examine more useful kinematic variables especially the $m_{\rm T2}$ variable in our analysis, which has been found can suppress most of background efficiently. In order to further improve the significance, we also apply two multivariate analysis approaches to optimise the signal and background discrimination.

In this work, we update the study explored in Ref. \cite{Baur:2002qd} and consider $3l 2j + \missE $ final state in more details and will focus on the sensitivity at the LHC and a future 100 TeV collider to the triple Higgs coupling. It is confirmed that the background from $Z W + \textrm{jets}$ is huge. In order to overcome this type of background, we deliberately consider the four three-lepton patterns: $e^\pm e^\pm \mu^\mp$ and $\mu^\pm \mu^\pm e^\mp$. Since it is essential to reconstruct the crucial information of signal events, we propose a partial reconstruction method and an efficient method to find the right combination of Higgs bosons. After the reconstruction, we further construct most of kinematical variables, especially the $m_{\rm T2}$ observable and examine their discrimination power to signal and background events. Considering the signal events are few, in order to enhance the significance, we apply two multivariate analysis methods to optimise the signal and background discrimination. Our results show that this channel can reach to a sensible significance i.e. 1.5 or so, at the $\SI{14}{TeV}$ LHC with $\SI{3000}{fb^{-1}}$ of integrated luminosity. We also extend the study to 100 TeV collisions with a luminosity 3 ab$^{-1}$ and find that the mode can be used to explore the discovery of all values of $\lambda_3$. When the SM will be confirmed, this mode can also be used to perform precision measure of $\lambda_3$ to the range [0.8, 1.5], which is comparable with the precision measurement by using the ratio of cross sections as pointed out in \cite{Goertz:2013kp}.

This work is organized as follows. We will describe the event generation of signal and background in Section II. Due to the existence of three neutrinos which consist of missing energy in events and are unable to be fully reconstructed, we will propose a partial reconstruction method for the visible objects and analyse the key kinematic features of signal events in the $3l 2j + \missE $ mode in Section II. By using the constructed kinematic observables, we will consider the sensitivity of 14 TeV LHC and a 100 TeV collider in Section IV. We will end this paper with discussions and conclusions.

\section{Event generation and Kinematic features of signal events}
We have generated the signal events in the following steps. 1) We have used the leading order matrix element computed by MadLoop/aMC@NLO\cite{Pittau:2012fn} and Gosam\cite{Cullen:2014yla}, which have taken into account the top-quark mass dependence in the loop evaluations. We have cross-checked the generated codes with the matrix element obtained by the FormCalc\cite{Hahn:1998yk}, and have found these independent approaches yielding the same results. 2) We perform the integration over  the whole phase space by using the VBFNLO code \cite{Arnold:2008rz,Arnold:2011wj,Baglio:2014uba} and obtain the total cross section.  3) After reaching a stable total cross section to the desired precision, we reweight each events in the phase space so as to yield the unweighted events at the parton-level.

For the LO cross sections, we used CTEQ6L1\cite{Pumplin:2002vw} PDF sets. We set the cuts in the phase space for the final Higgs bosons as $|\eta(H)| < 5$ and $P_t(H)>1$ GeV. We set the renormalisation and factorisation scales as $\mu_r=\mu_f=\sqrt{\hat{s}}$, and have reproduce the LO total cross section as $22.8$ fb, which agrees with our previous results \cite{Li:2013flc}. 

Using the unweighted events, we use the package DECAY provided in MadGraph5\cite{Alwall:2011uj} to decay Higgs into a pair of W bosons (one is on-shell and the other is off-shell) and further to decay W bosons into quarks and leptons. Therefore, all spin correlation information in the final states have been taken into account in the data sample. Before considering lepton and jet and missing energy reconstruction at detector level, we use PYTHIA6\cite{Sjostrand:2006za} to perform parton showering. 

For the background processes, we use Madgraph5\cite{Alwall:2011uj} to generate events, and also shower it with PYTHIA6\cite{Sjostrand:2006za}. In order to avoid the double-counting issue of jets originated from matrix element calculation and the parton shower, we apply the MLM-matching implemented in Madgraph5\cite{Alwall:2011uj}. In practice, for the background events of $ t \bar{\textrm{t}} W$, we include both processes $p p \rightarrow t \bar t W $ and $p p \rightarrow t \bar t W + j$ to form an inclusive dataset. For the background events of $ W W W$, we include processes $p p \rightarrow W W W $ and $p p \rightarrow W W W + j$ and $p p \rightarrow W W W + j j$, and similarly for ZW,HW backgrounds.
The $ZW^{\pm}\to l^{+}l^{-}W$ background are generated by using exact matrix element which including off-shell Z and $\gamma$ effects, and other backgrounds are all generated on mass-shell. We ignore $ZZ$ background due to it require one lepton is missing and should be much smaller than $ZW$ background. The background $t\bar{t}Z$,$ZWW$ is also ignored because it is much(20~30) smaller compared to the $t\bar{\textrm{t}}W$,$WWW$ background, correspondingly. We also ignore $t\bar{t}t\bar{t}$ due to its tiny cross section and the efficient rejection by b-taggings. We would like to mention that in all background event generation that the decay correlation for all final states have been correctly accounted for.

For the analysis at the detector level, we first reconstruct isolated leptons in each event. After that  we pass all the rest of visible particles to FastJet\cite{oai:arXiv.org:1111.6097} to cluster to into jets. We adopt the anti-kt algorithm \cite{Cacciari:2008gp} with cone parameter $R=0.4$. After that, the transverse missing energy is reconstructed. In this study, we have neither taken into account the magnetic effects for charged tracks nor the energy smearing effects for leptons and jets. Therefore, our analysis should be regarded as a hadron level analysis.

\begin{figure*}[!htbp]
\centering
\subfigure[~Pt of 1st lepton
\label{fig:l1pt}]
{\includegraphics[width=.45\textwidth]{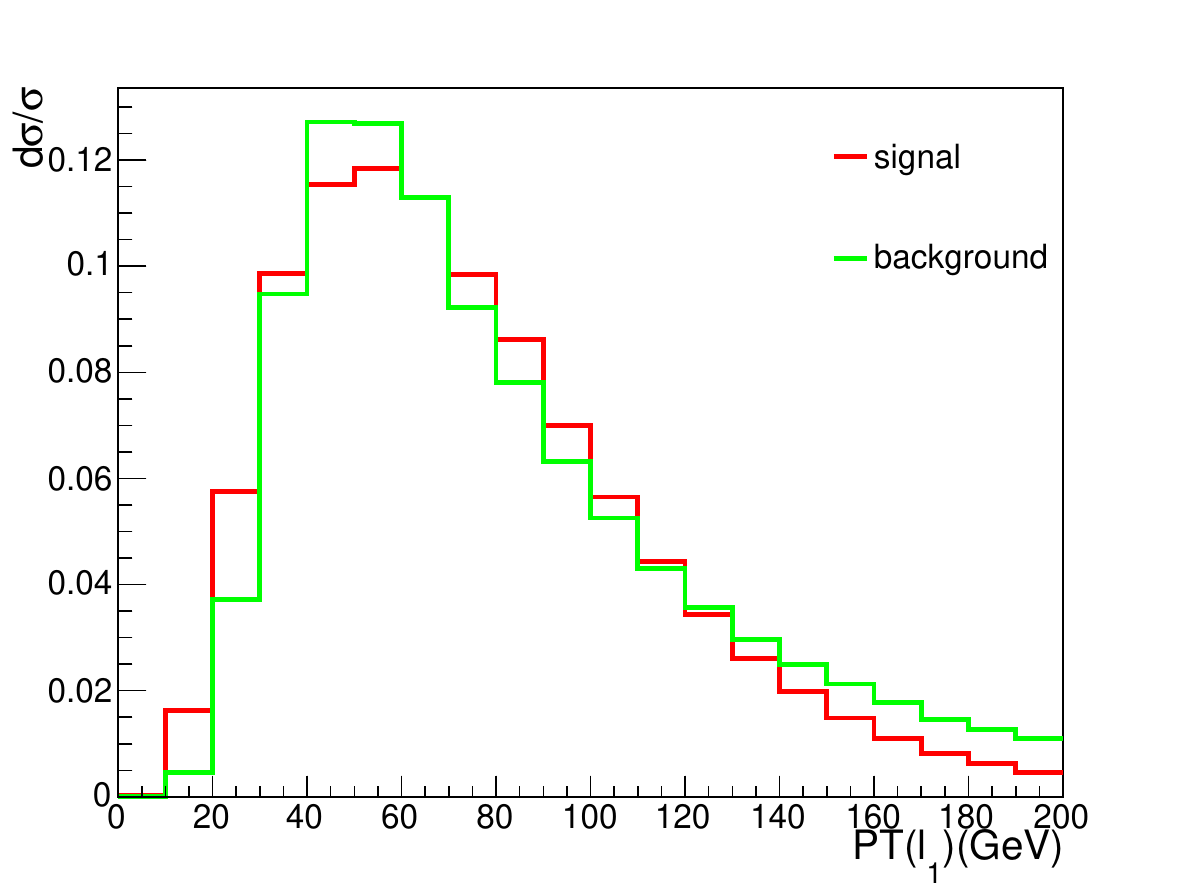}}
\subfigure[~Pt of 2nd lepton
\label{fig:l2pt}]
{\includegraphics[width=.45\textwidth]{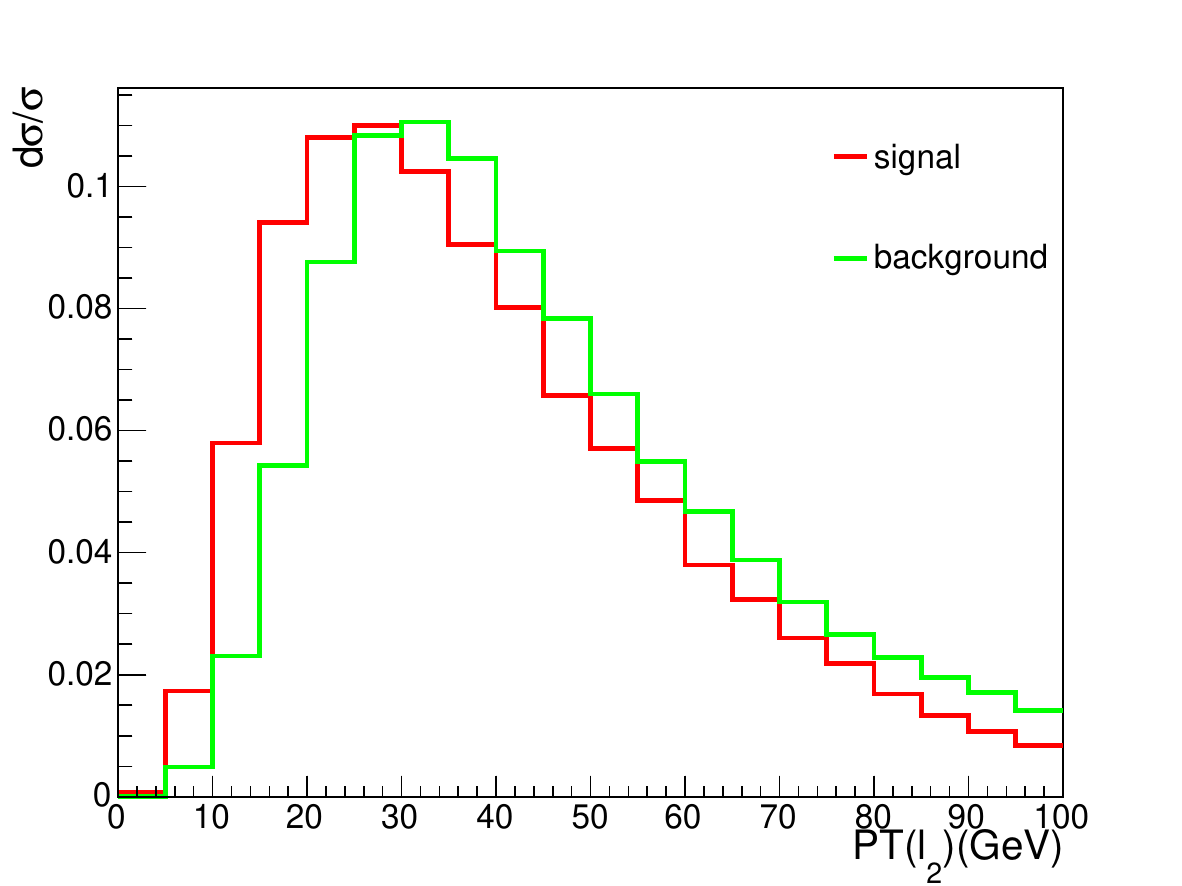}} \\
\subfigure[~Pt of 3rd lepton
\label{fig:l3pt}] 
{\includegraphics[width=.45\textwidth]{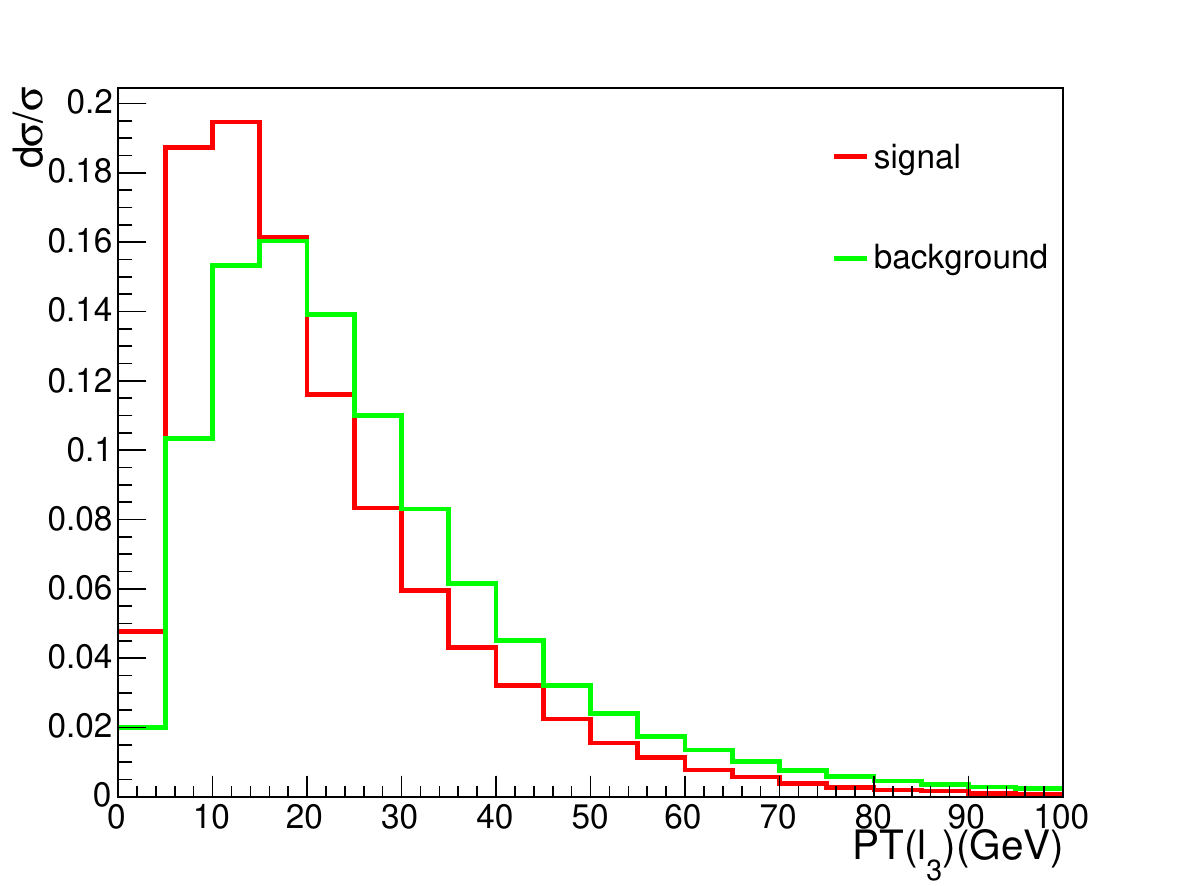}} 
\subfigure[~Pt of 1st Jet
\label{fig:j1pt}]
{\includegraphics[width=.45\textwidth]{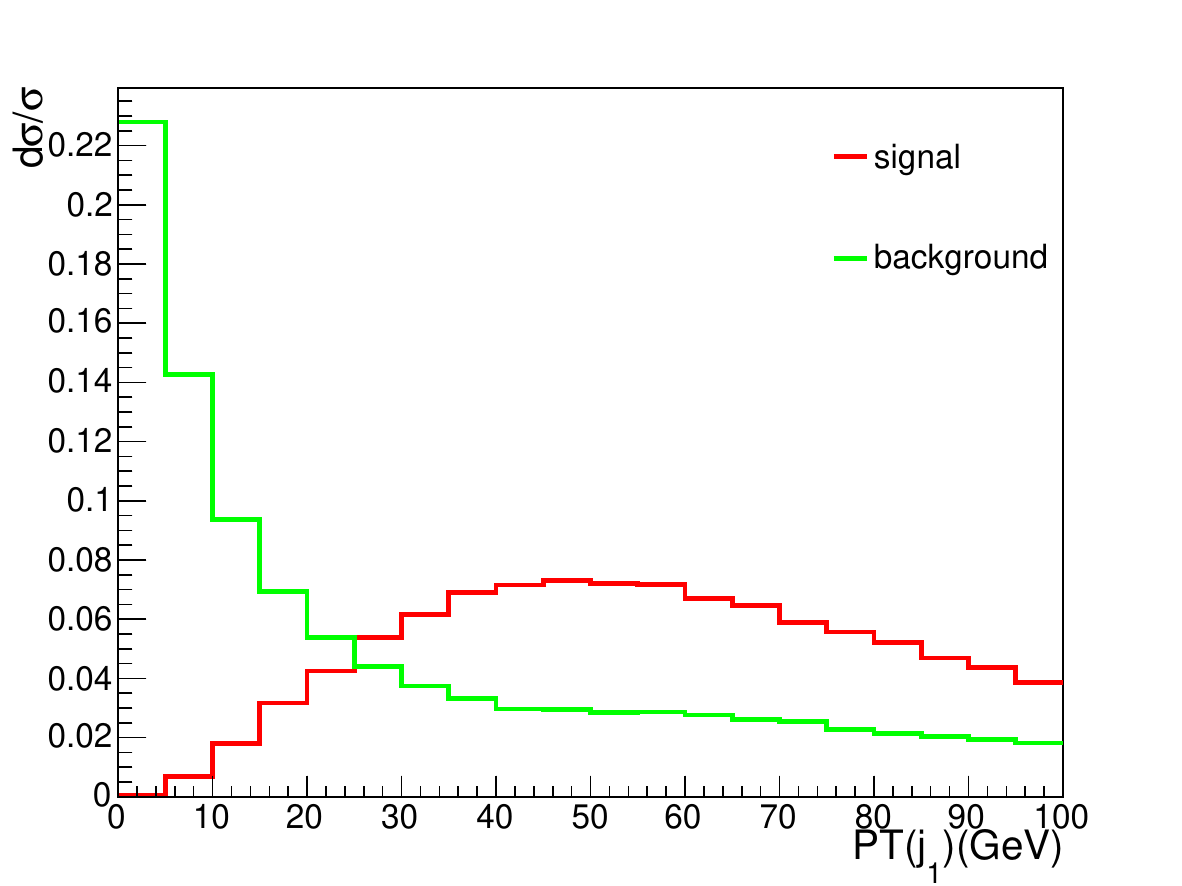}} \\
\subfigure[~Pt of 2nd Jet
\label{fig:j2pt}] 
{\includegraphics[width=.45\textwidth]{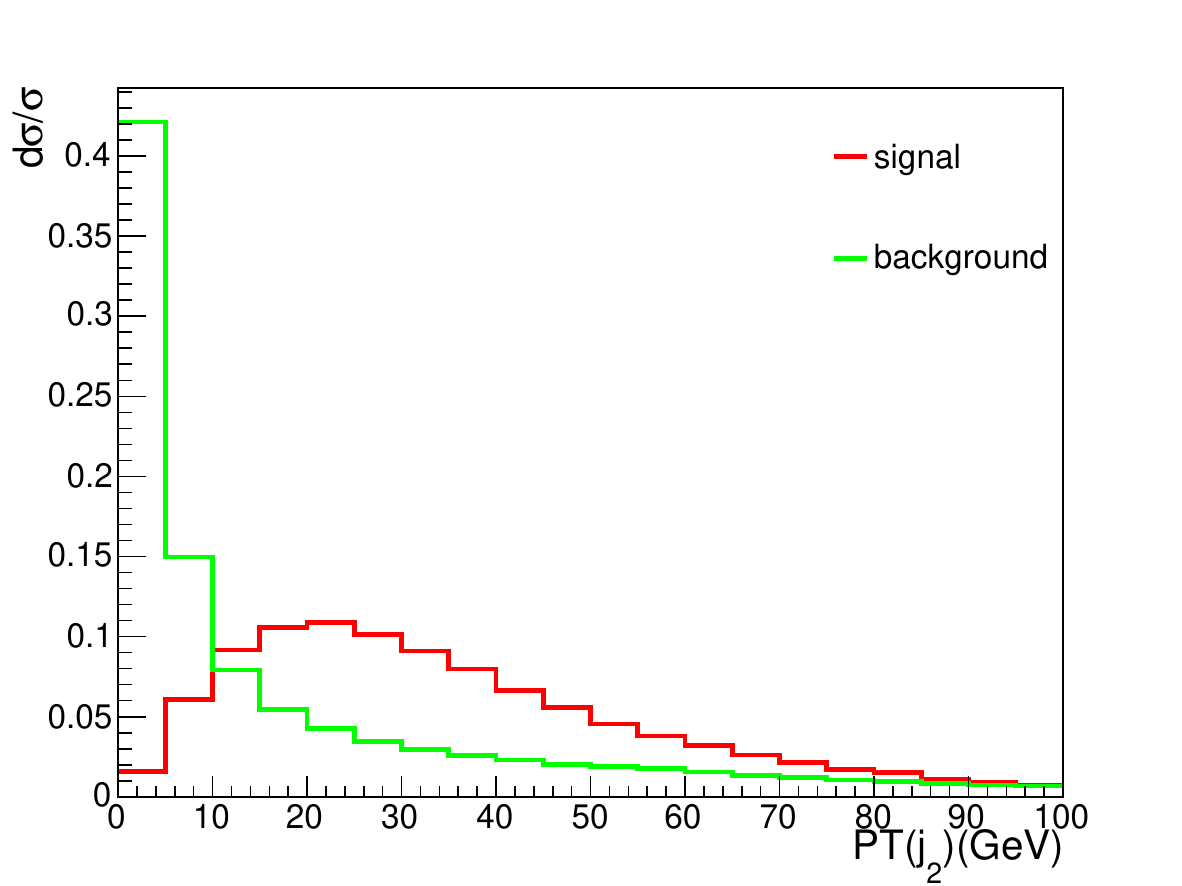}}
\subfigure[~Invariant mass of two leading jets
\label{fig:m2j}]
{\includegraphics[width=.45\textwidth]{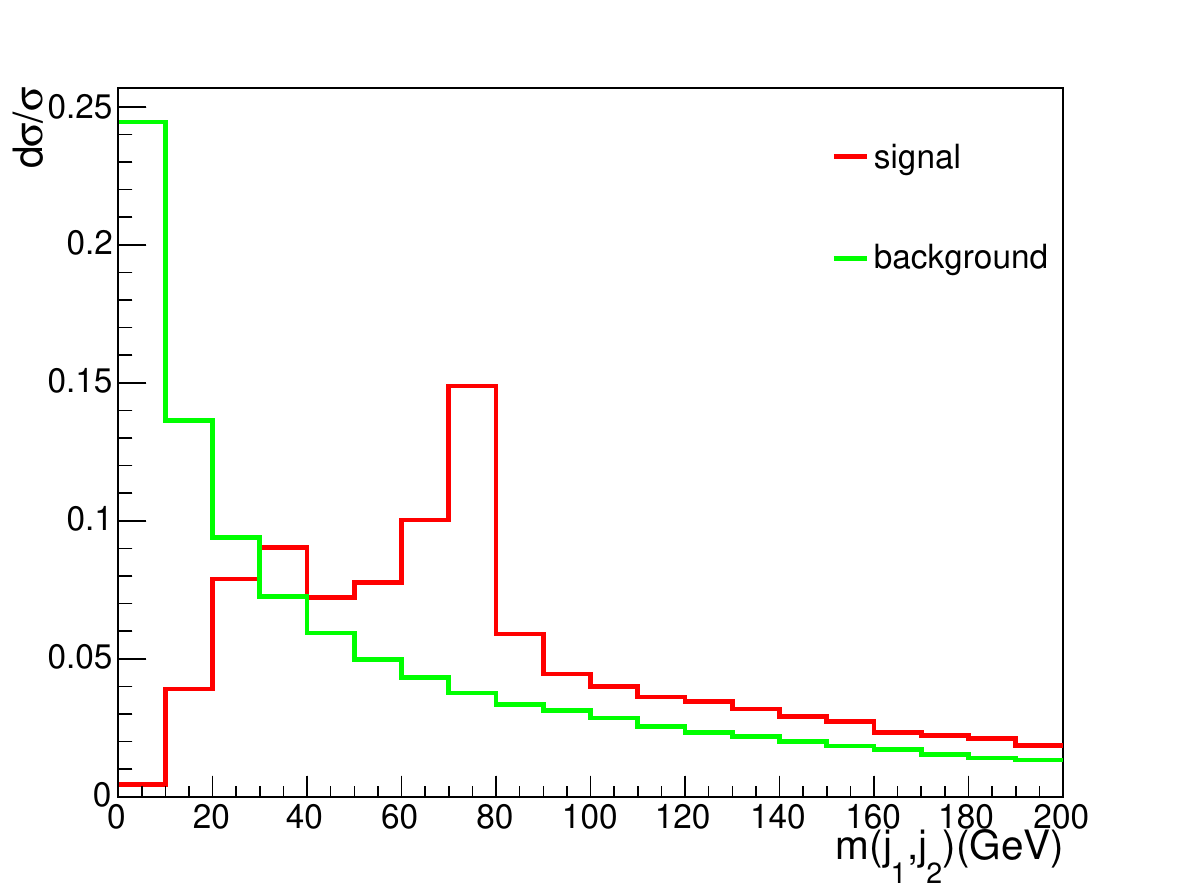}}
\caption{The transverse momentum of leading three leptons and leading two jets and the invariant mass of two jets as well are shown at parton level. Both signal and background are normalized one.}
\label{fig:constr}
\end{figure*}

In order to suppress the dominant background and select the most relevant events, we introduce all the following pre-selection cuts at event-by-event level:
\begin{itemize}
\item We veto events with isolated and energetic photon(s) with $P_t(\gamma) > 10 $ GeV and $|\eta(\gamma)| < 2.5$;
\item In order to suppress the large background from $t\bar{t} W$ and $t \bar{t} H$, we veto events with tagged B jets. In our simulation, the tagging efficiency of B jets is assumed $60\%$. Therefore, roughly the background from $t\bar{t} W$ and $t\bar{t} H$ can be suppressed by a factor $0.16$.
\item The preselection rule for three isolated leptons is found to be crucial. We demand that there are exactly three isolated leptons being found, with the requirement that the first leading lepton should have a momentum larger than $30$ GeV, the next leading lepton should larger than 10 GeV,  the softest lepton should larger than $10$ GeV. Since there must be a lepton coming from the on-shell W boson decay, so we require the leading lepton should be hard enough. At the meantime, there must be a  softer lepton which comes from off-shell $W$ bosons decay. In order to increase the acceptance for signal, we deliberately lower the momentum of the third lepton. In Figure (\ref{fig:constr}a-c), we present the distributions of 3 leptons. Considering that threshold of lepton reconstruction is around $3$ GeV and if lepton is larger than 5-8 GeV at the both CMS and ATLAS detectors lepton reconstruction efficiency can be $95\%$ \cite{atdr,ctdr}, to find the soft leptons with $P_t  > 10$ GeV in the signal event should be plausible. 
\item In order to suppress the large background $Z/\gamma W$+jj, we only consider the following four modes with two leptons of same sign and same flavor plus an extra different flavored lepton:  $e^- e^- \mu^+$, $e^+ e^+ \mu^-$, $\mu^+ \mu^+ e^-$, and $\mu^- \mu^- e^+$. After this preselection cut, we noticed that the background events from the processes $Z/\gamma W$+jj can be safely neglected.
\item At least two jets in the events are required to be successfully reconstructed, i.e. $n_j\geq 2$ and $|\eta(j)| <2.5$. Among those reconstructed jets, there are two jets which could come from a W boson either on-shell or off-shell. In order to increase the acceptance of signal, we only consider those jets with transverse momentum larger than 15 GeV. We show the distributions of these two leading jet in Figure  (\ref{fig:constr}d-e). We also show the invariant mass of these two jets in Figure. (\ref{fig:constr}f). It is noticed that the invariant mass of in signal events can produce two peaks, one is near the value of $M_W$ and the other is near that of $M_H - M_W$.
\item The missing transverse momentum is required to be larger than $\missE_T > 20$ GeV due to neutrinos in the signal processes. The requirement on large missing energy is also useful in order to suppress the huge QCD processes and to save the computing time.
\end{itemize}

\begin{table}[th]
\begin{center}%
\begin{tabular}{|c|c|c|c|}
\hline
processes &  $\sigma^{\textrm{LO}}$ $\times$ branching fraction (fb) & K factors & No. Events after preselection cuts \\ \hline
Signal $gg\to H H $ & 3.0 $\times 10^{-2} $  & 1.8 \cite{deFlorian:2013jea} & 16.3 \\ \hline
$H W^\pm$ & 1.2 & 1.2 \cite{Ferrera:2011bk} & 119.4 \\ \hline
$WWW$ & 1.4& 1.8 \cite{Binoth:2008kt} & 363.9 \\\hline
$\textrm{t} \bar{\textrm{t}} W^\pm$ & 4.6 & 1.3 \cite{Campbell:2012dh} & 451.4 \\ \hline
$\textrm{t}\bar{\textrm{t}}H$ & 2.1 & 1.2  \cite{Beenakker:2002nc} & 101.3   \\\hline
$Z W, \gamma W$ & 233 & 1.8 \cite{Campbell:1999ah} & $\sim$ 0  \\\hline
$S/B$ &  \multicolumn{3}{c|} {0.02} \\\hline \hline
$S/\sqrt{B} $ & \multicolumn{3}{c|} {0.53}  \\\hline
\end{tabular}
\end{center}
\caption{The number of signal and background events are shown. Here we assume the total integrated luminosity as 3 ab${^{-1}}$. \label{table1No}}%
\end{table}

The LHC detectors can record signal events, which can be triggered by both energetic charged lepton and large missing energy. From Table \ref{table1No}, we observe that the number of background events is around 200 times larger than that of signal events, and it is indeed a challenge if we want  to distinguish the signal and the background successfully .


In order to distinguish the signal and background event, we have to resort to the reconstruction procedure so as to extract the most important information of signal. Since the Higgs boson is a neutral particle, for the decay mode $\ell_1^\pm \ell_2^\pm \ell_3^\mp 2j+\sl{E}$, without considering the neutrinos, there are only two possible combinations for a pair of Higgs boson decay: ($H(\ell_1^\pm \ell_3^\mp)$, $H(\ell_2^\pm jj)$) or  ($H(\ell_2^\pm \ell_3^\mp), H(\ell_1^\pm jj))$. For the convenience of later study, hereby we label the first Higgs boson as leptonic one ($H(\ell\ell)$) and the second one as semileptonic one ($H(\ell jj)$). 

\begin{figure}[!htb]
\begin{center}
\includegraphics[width=0.9 \columnwidth]{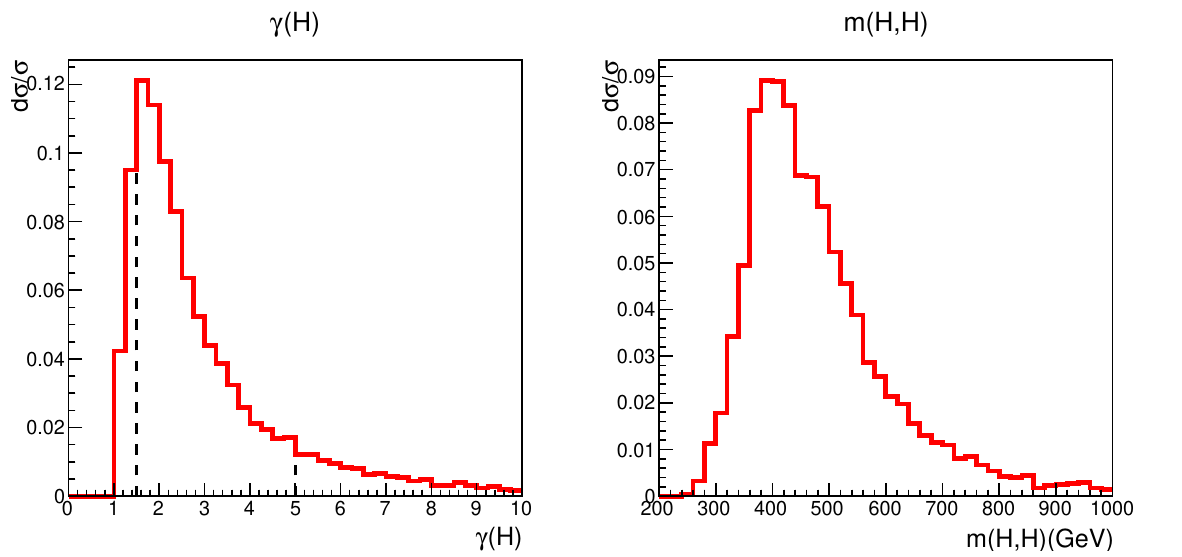}
\caption{The $\gamma$ factor ($\gamma=E(H)/m_H$) and the invariant mass of Higgs pair are shown at the parton level. }
\label{partonh}
\end{center}
\end{figure}

As we can read out from the left panel Fig.\ref{partonh}, each of Higgs bosons is moderately boosted when produced and the peak value of $\gamma$ (here $\gamma$ is defined as $E(H)/m_H$, which is a measure to the boost) is around 2. The fraction of highly boosted Higgs boson $\gamma > 5 $ is around $13\%$ or so, while the fraction of moderately boosted Higgs boson $\gamma > 1.5$ is around $87\%$. In the right panel of Fig.\ref{partonh}, we show the invariant mass of Higgs pair. It is observed that the peak region of the invariant mass of Higgs pair is around $360-540$ GeV, which explains why most of the Higgs pairs are boosted.

\section{Reconstruction of Signal and Observables}
To know the right combination is crucial to reconstruct the kinematic features of the signal and can provide important information to suppress background events. For that purpose, we need find a way to determine the right combination reliably.

\subsection{Determination of the right combination}
Fortunately, the problem at hand is not complicated after the preselection and the number of combinations is not formidable. It is observed that in the selected events, there are three leptons in total. Two leptons with same sign and same flavour must come from different Higgs bosons, there are only two possible combinations for each a signal event. The remaining task is to find the right combination by exploiting the kinematics of Higgs bosons.

\begin{table}[th]
\begin{center}
\begin{tabular}{|c|c|}
\hline
Methods &  The percentage of correctness ($\%$) \\ \hline
$|m_{H(ll)}-m_{H(ljj)}|$ &  68.9   \\ \hline
$\Delta R(l^{\pm}, l^{\mp})$ &   85.0 \\ \hline
 $\Delta R(l^{\pm}, W_{jj})$ &  89.9 \\ \hline
 $P_t[H(ll)]+P_t[H(ljj)]$ &  90.3 \\ \hline
 $\Delta_R(H(ll),H(ljj))$ &  92.0 \\ \hline
 $m_{H(ll)}+m_{H(ljj)}$  &  95.4 \\ \hline
\end{tabular}
\end{center}
\caption{The principal observables to choose the right combination in six methods and the percentage of correctness at the parton level are tabulated.}
\label{tableoptions}
\end{table}

Keeping those kinematic features of Higgs bosons in signal events exposed in the last section, we consider the following six individual methods by using different observables to pick out the combination from two as a solution for each event. In Table 
\ref{tableoptions}, we tabulate the principal observables and the percentage of correctness to pick out the right combinations at parton level, which serves as an important guide for our later analysis at hadronic/detector level. Below, we examine the efficiency of these six methods one by one.

\begin{itemize}
\item[1] In the first method, we utilize the fact that the mass of the Higgs bosons in the pair production must be the same. But due to the missing energy carrying away by the neutrinos, if we use the condition that the mass difference should be smaller we find that we can only reach the right combination in $69\%$.
\item[2] In the second method, we use the fact that most of Higgs bosons are moderately boosted and two leptons from the leptonic Higgs are tended to be close in spatial separation due to the spin correlation of W bosons from Higgs decay\cite{Cao:2007du,Alwall:2007st}, therefore two leptons from it decay should have a smaller angle separation $\Delta R(\ell^+, \ell^-)$. We notice that by using this observable, the right combination can be determined by  $85\%$.
\item[3] In the third method, we use the semileptonic Higgs as a guide by requiring the smaller angle separation $\Delta R(\ell, W_{jj})$ between a lepton and a hadronic W. Due to the smaller energy loss from its decay, we observe a higher percentage in choosing the right combination when compared with the second method, which can reach to $90\%$.
\item[4] In the forth method, we resort to the scalar sum of the transverse momenta of Higgs pair (without taking into account the missing transverse momentum), which should be large due to the energy conservation in the transverse direction. When the wrong combination is made, the scalar sum is found decrease. The method can have similar performance as the third method.
\item[5] In the fifth method, we exploit the fact that two Higgs bosons mostly fly back-to-back in 3d space, therefore the angular separation of them $\Delta R(H(\ell \ell),H(\ell j j))=\sqrt{\Delta \phi^2 + \Delta \eta^2} $  should be large. For the two possible combinations, we choose the one which yields the larger $\Delta R(H(\ell \ell), H(\ell jj ))$ as the solution and we observe that this method can arrive at an efficiency $92\%$. 
\item[6] In the sixth method, we compute the invariant mass of each Higgs boson from visible objects, which can be labelled as $m_{H(\ell\ell)}$ and $m_{H(\ell j j)}$, respectively.  Then we sum these two masses $m_{H(\ell\ell)}+m_{H(\ell j j)}$. We choose the combination which yield a smaller value as the solution. We notice that this method reach the highest percentage of correction combination up to $95\%$.
\end{itemize}

Therefore, in the following analysis, we will use the sum of invariant masses of Higgs bosons, i.e. the sixth method, to determine the combination and extract the relevant experimental observables at hadron/detector level. 

Another remarkable aspect is the missing energy, or more precisely the missing transverse momentum. In signal events, there are three neutrinos in total, which should have 9 degree of freedom to determine the full phase space. But we can obtain have at most 5 constraints. So in principle, it is impossible to solve the kinematics at event-by-event level. Nevertheless, in the hypotheses of pair production, we can split the transverse missing momenta of each event into two parts. The first part will combine with the lepton pair to reconstruct the transverse mass of the first Higgs boson, and the second part should combine with the rest of objects in the event to form that of the second Higgs boson. Below we will explore the variable $m_{T2}$.

\subsection{The variable $m_{T2}$}

The variable, $m_T$, the transverse mass  of W boson has played a crucial role for the discovery of W boson\cite{Smith:1983aa}. The extended variable $m_{T2}$ was introduced to extract the information of particle mass in pair production processes at hadron colliders \cite{Lester:1999tx,Barr:2003rg} when the information of both the mass and longitudinal components of invisible particles are missing.

The original setup assumed the production of a pair of particles $A_1$ and $A_2$, then particles $A_i$ decay into invisible particle $B_i$ and visible particle $C_i$, for example: $pp\to A_1A_2,A_1\to B_1C_1,A_2\to B_2C_2$, where $B_i$ particles denote invisible particles like neutrinos and neutralinos of the SUSY and $C_i$ particles denote visible particles like leptons or jets of which the energies and momenta can be reconstructed by detectors. At hardon colliders, only the sum of transverse momentum of $B_1$ and $B_2$ which is denoted as $\missE_T$ can be reconstructed by assuming the energy conservation in the transverse directions. In experiments, the missing transverse momenta $\missE_T$ can be reconstructed by using the particle flow algorithm, for instance. Since the energy and longitudinal components are missing and in principle it is impossible to reconstruct the mass of $m_A$, but we can define the transverse mass of particle A from the transverse momenta the particles B and C: $M_T^2(P_T(B),P_T(C))=(E_T(B)+E_T(C))^2-(\vec{P}_T(B)+\vec{P}_T(C))^2$, where the transverse energy is defined as $E_T^2=\vec{P}_T^2+M^2$. There exists an inequality $M_T(B,C)\leq M(A)$. 

In practice, to construct the variable $m_{T2}$ we split the missing transverse momenta into two parts and to find the minimal of the maximal of reconstructed transverse mass:
\begin{gather*}
m_{T2}=\min_{\vec{P}_{T1}+\vec{P}_{T2}=\vec{\missE_T}}\left\{\max\left[m_T^2(\vec{P}_T(B_1),\vec{P}_{T}(C_1)),m_T^2(\vec{P}_T(B_2),\vec{P}_{T}(C_2))\right]\right\}\,.
\end{gather*}
For each an event, the minimization is taken over all possible transverse momentum splitting. For a pair production event, the $m_{T2}$ corresponds to find the solution where both the reconstructed transverse masses from each decay chain are equal. Recently, there are more studies on the $m_{T2}$ variable and its variants, interested readers can refer \cite{mt2review,Barr:2011xt,Cho:2014naa} for more information.

Obviously, this variable can be generalized to the cases where either particles B or C are not a single particle then either $A_1$ or $A_2$ can decay into different final states. In the case at our hand, the leptonic Higgs boson decays into $(l^{\pm}l^{\mp})(\nu\nu)$, and the semi-leptonic Higgs decays into $(l^{\pm}jj)(\nu)$. In the invisible part of the leptonic Higgs contains two neutrinos and their invariant mass is unable to know. Considering that the variable $m_{T2}$ is a monotonous increasing function on the $m_{12}$, for simplicity, we choose it as zero.

\begin{figure*}[!htbp]
\centering
\subfigure[~Mass of leptonic Higgs
\label{fig:mhll}]
{\includegraphics[width=.45 \textwidth]{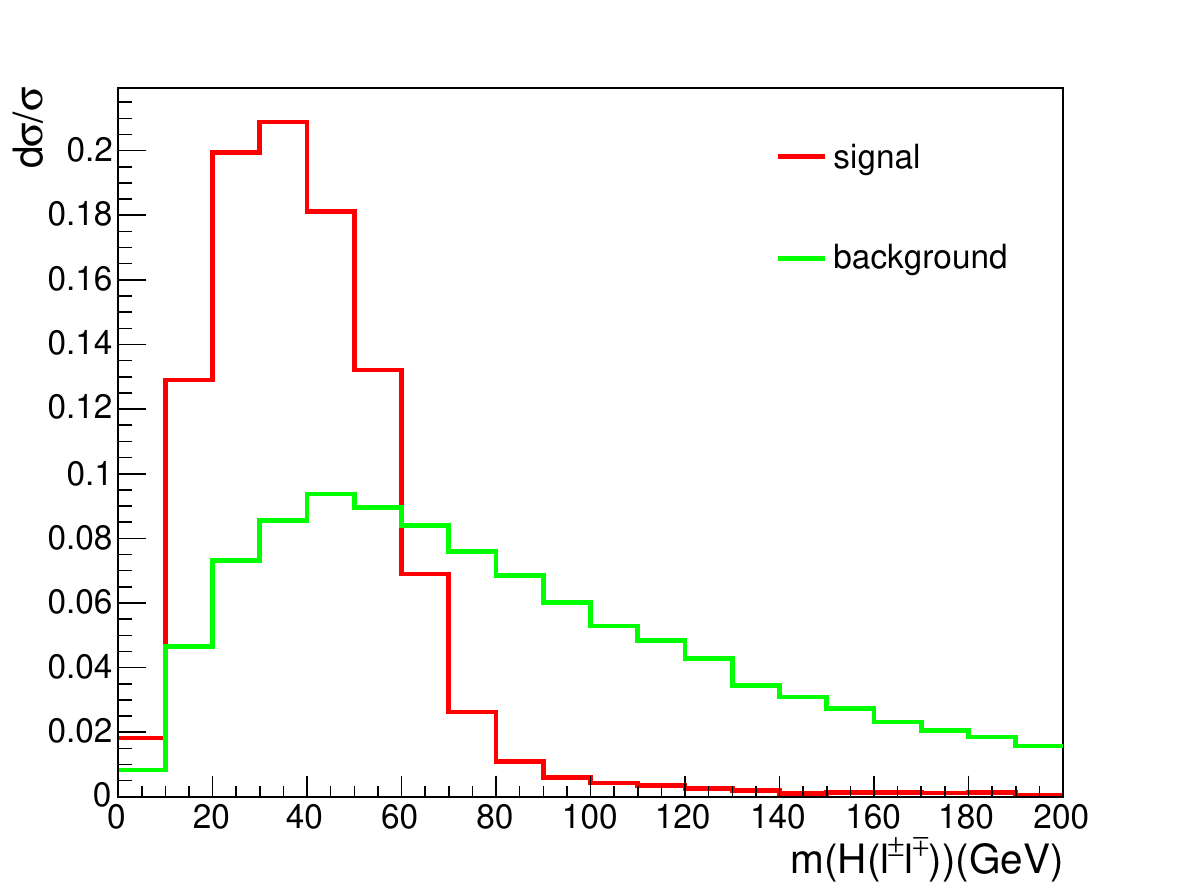}}
\subfigure[~Mass of semileptonic Higgs
\label{fig:mhlj}]
{\includegraphics[width=.45\textwidth]{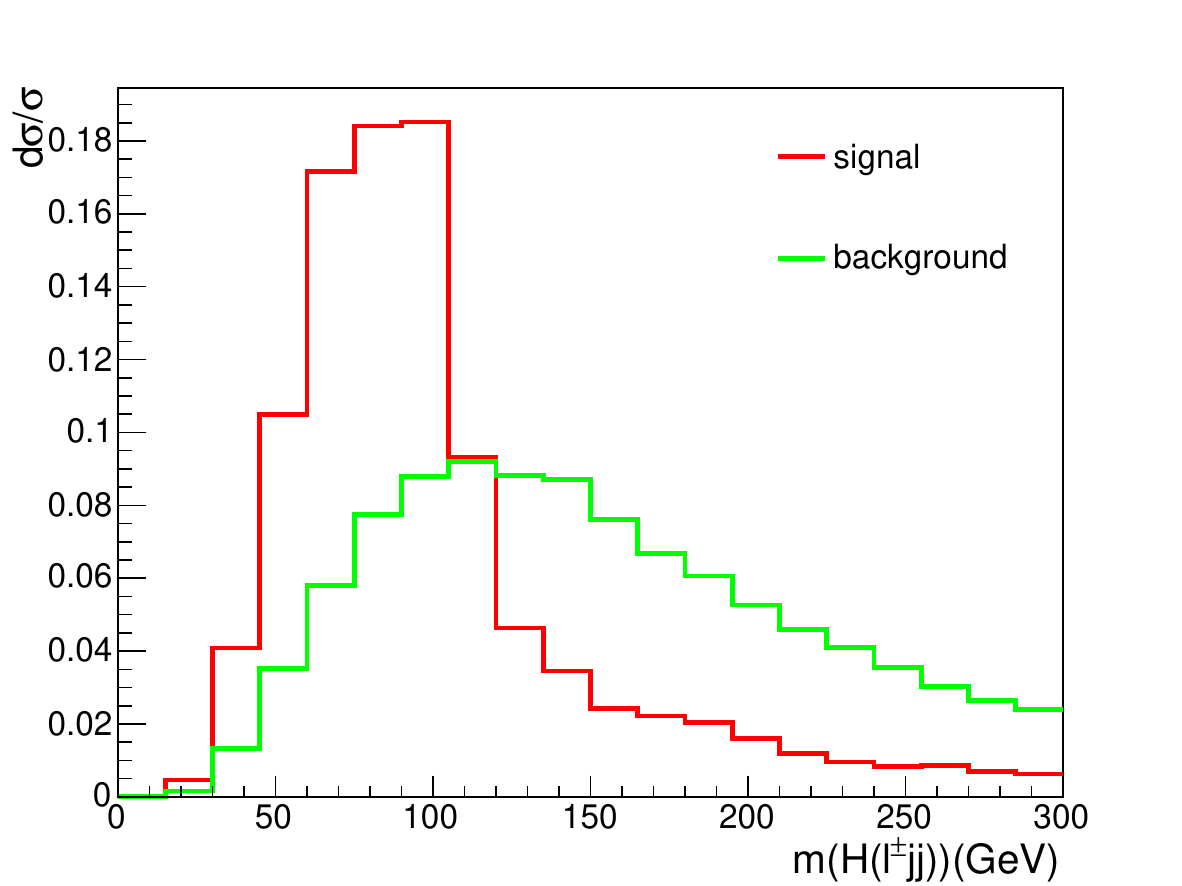}} \\
\subfigure[~ The reconstructed W/$W^*$ mass
\label{fig:drlw}]
{\includegraphics[width=.45\textwidth]{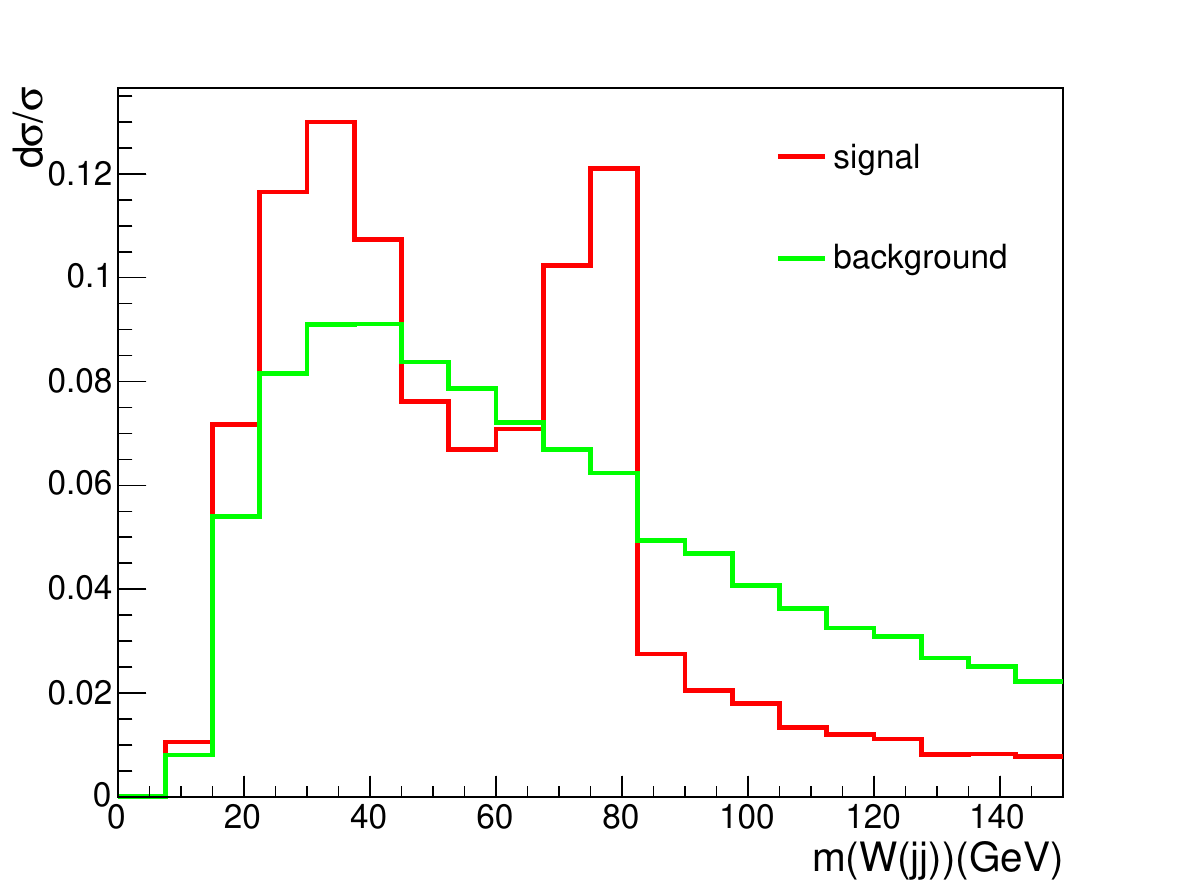}} 
\subfigure[~ $m_{T2}$
\label{fig:mt2}]
{\includegraphics[width=.45\textwidth]{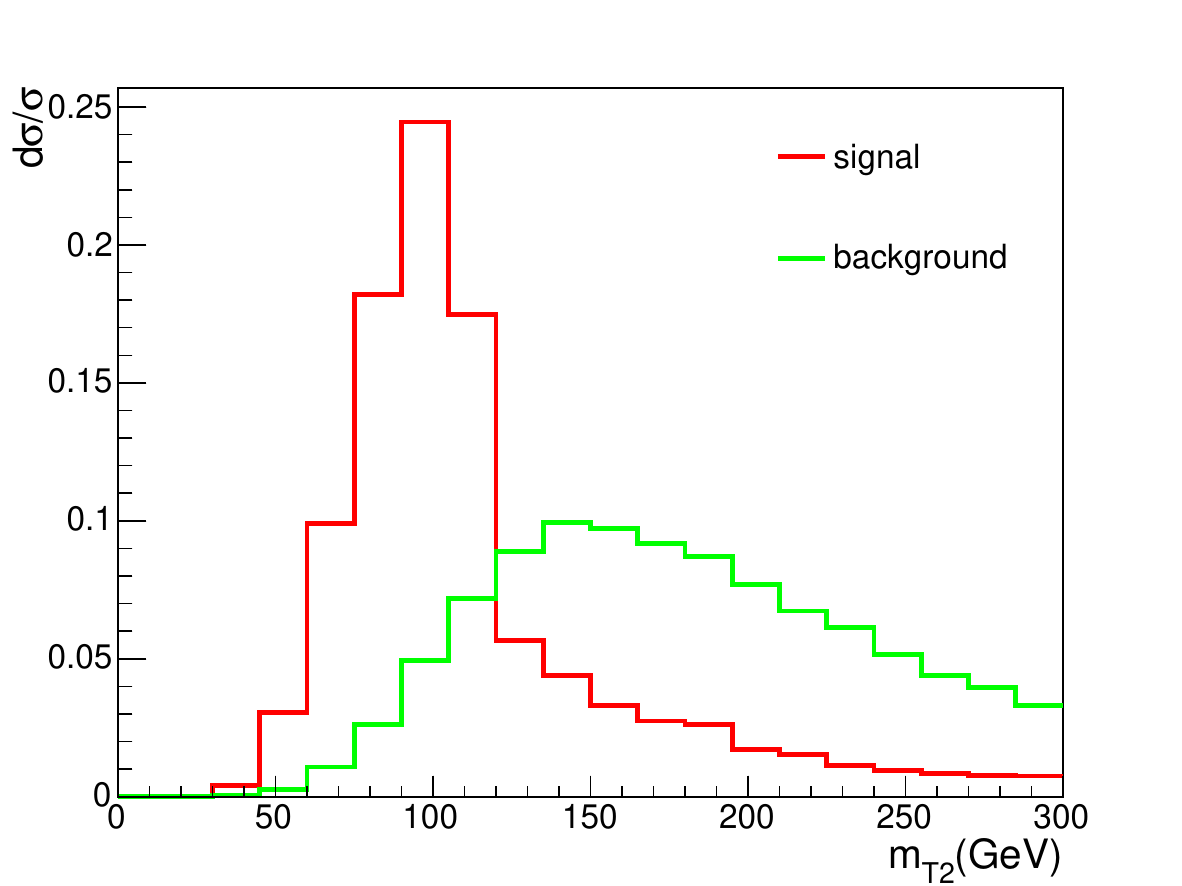}}
\caption{Four crucial reconstructed kinematic observables at hadron level are demonstrated.}
\label{fig:recon}
\end{figure*}

For the case at our hand, after the splitting of missing transverse momenta, we can construct the transverse mass of Higgs boson by using the $m_{T2}$ code \cite{mt2code}. So the first part of the split $\missE_T$ should correspond to the combination of two neutrinos, and the second part of $\missE_T$ should correspond to a neutrino.  So that the transverse mass of Higgs boson can be constructed. The quantity is called as the $m_{T2}$ variable, which has utilised information of both the visible and invisible objects in an event.  It is remarkable that this quantity is the most sensitive observable to distinguish signal and background, as shown in both Fig. (\ref{fig:recon}d) and Table (\ref{table:eachcut}).

In Figure (\ref{fig:recon}), we show the line shapes of signal and background events in terms of $m_{H(\ell\ell)}$, $\Delta R(\ell, jj)$, $m_{H(\ell jj)}$, and $m_{T2}$. From the line shapes, we introduce a one-dimensional cut for each of these observables. In Table (\ref{table:eachcut}), we tabulate the efficiency of each cut.
\begin{table}[th]
\begin{center}%
\begin{tabular}
[c]{|c|c|c|c|c|c|c|c|}\hline
 & signal H H &  t $\bar{\textrm{t}}$ + W  & H W & W W W & t $\bar{\textrm{t}}$  H  & S/B & $S/\sqrt{B}$\\ \hline \hline
No. after preselection  &   $13.7$ &$317.2$ & $94.4$ & $400.6$ & $101.3$ & $1.5 \times 10^{-2} $ & $0.45$ \\ \hline \hline
$m_{jj} < 80$ GeV        & $10.6$ & $153.7$ & $53.3$ & $189.6$ & $78.7$ & $2.2 \times 10^{-2}$ & $0.49$ \\ \hline \hline
$m_{H(\ell,jj)} < 110$ GeV &  $9.6$ &   $70.6$ & $27.8$ & $78.2$ & $54.8$ & $4.2 \times 10^{-2} $ & $0.63$ \\ \hline \hline
$m_{H(\ell \ell)} < 55$ GeV  &  $11.2$ & $76.9$ &   $65.0$ & $92.8$ & $53.6$ & $3.9 \times 10^{-2}$ & $0.66$ \\ \hline \hline
$m_{T2} < 110$ GeV &  $8.4$ &   $18.4$ & $16.7$ & $19.1$ & $27.1$ & $1.0 \times 10^{-1}$ & 0.93 \\ \hline \hline
\end{tabular}
\end{center}
\caption{The efficiency of four crucial cuts are demonstrated. To appreciate the efficiency of each cut, we also provide the values of $S/B$ and $S/\sqrt{B}$ after each a cut.}%
\label{table:eachcut}%
\end{table}
It is noticed the observable $m_{T2}$ can have the best distinguishing power and the observable $m_{H(\ell\ell)}$ is the second powerful observable. From Table (\ref{table:eachcut}), it is remarkable that the backgrounds from $t \bar{\textrm{t}} W$ and $WWW$ can be heavily affected by this quantity since they are not pair production processes in nature. While for the process $H W$, extra jets from initial state radiation should be used to balance the pair production hypothesis.

\begin{table}[th]
\begin{center}%
\begin{tabular}
[c]{|c|c|c|c|c|c|}\hline
 & signal H H &  t $\bar{\textrm{t}}$ + W  & H W & W W W & t $\bar{\textrm{t}}$ H  \\ \hline \hline
No. after preselection  &   $13.7$ &$317.2$ & $94.4$ & $400.6$ & $101.3$ \\ \hline \hline
$m_{T2} < 110$ GeV &  $8.4$ &   $18.4$ & $16.7$ & $19.1$ & $27.1$ \\ \hline \hline
$m_{H(\ell \ell)} < 55$ GeV & $7.2$ & $10.5$ & $13.0$ & $11.5$ & $19.0$ \\ \hline \hline
No. of jets $<= 4$ & $6.2$ & $8.0$ & $12.0$ & $8.8$ & $7.9$ \\ \hline \hline
$S/B$ &  \multicolumn{5}{c|} {0.17} \\ \hline
$S/\sqrt{B} $ & \multicolumn{5}{c|} {1.0}  \\ \hline
\end{tabular}
\end{center}
\caption{The effects of each cut in the cut-based method are demonstrated in a sequential way. After all cuts, the values of S/B and $S/\sqrt{B}$ are provided. }%
\label{table:cbm}%
\end{table}

When we combine all of these cuts into a cut-based method, we arrive at the significance given in Table (\ref{table:cbm}). After using the quantities extracted from our reconstruction procedure, we notice that the $S/B$ can be improved by a factor $10$ or so. Compared with the results given in Table (\ref{table1No}), we observe a big gain in both $S/B$ and $S/\sqrt{B}$. The gain is mainly yielded by the success of suppression to the background processes $t \bar{\textrm{t}} W$ and $W W W$. In contrast, the suppression to the background $H W$ is relatively limited due to the appearance of a real Higgs boson in the process and our reconstruction procedure can find the Higgs bosons in the events. For example, the cut $m_{H(\ell \ell)}<55$ GeV has no serious effects to this background process. But, instead, the variables from semi-leptonic Higgs can impose a significant suppression to this type of background.
\begin{table}[th]
\begin{center}%
\begin{tabular}
[c]{|c|c|c|c|c|}\hline
 & after preselection cuts &  Cut-based method & MLP method & BDT method  \\ 
 & & & $N_{NN} > 0.82 $ & $N_{BDT} > 0.41$\\ \hline
No. of Signal& $13.7$ & $6.2$ & $5.7$ & $3.8$ \\ \hline
No. of Background & $913.5$ & $36.8$ & $21.7$ & $6.2$ \\ \hline
$S/B$ &  $1.5\times10^{-2}$& $1.7 \times 10^{-1}$ & $2.6 \times 10^{-1}$ & $6.2\times 10^{-1}$ \\ \hline
$S/\sqrt{B}$ &  $0.45$& $1.0$ & $1.2$ & $1.5$ \\ \hline
\end{tabular}
\end{center}
\caption{Comparison of significance among three analysis methods are shown.}
\label{table:TMVA1.0}%
\end{table}

\section{The sensitivity to triple Higgs coupling}

\subsection{The sensitivity to $\lambda_3$ at LHC 14 TeV with a 3 ab$^{-1}$ dataset}

\begin{figure*}[!htbp]
\centering
\subfigure[~Pt Sum of objects used for reconstruciton
\label{fig:j1pt}]
{\includegraphics[width=.45\textwidth]{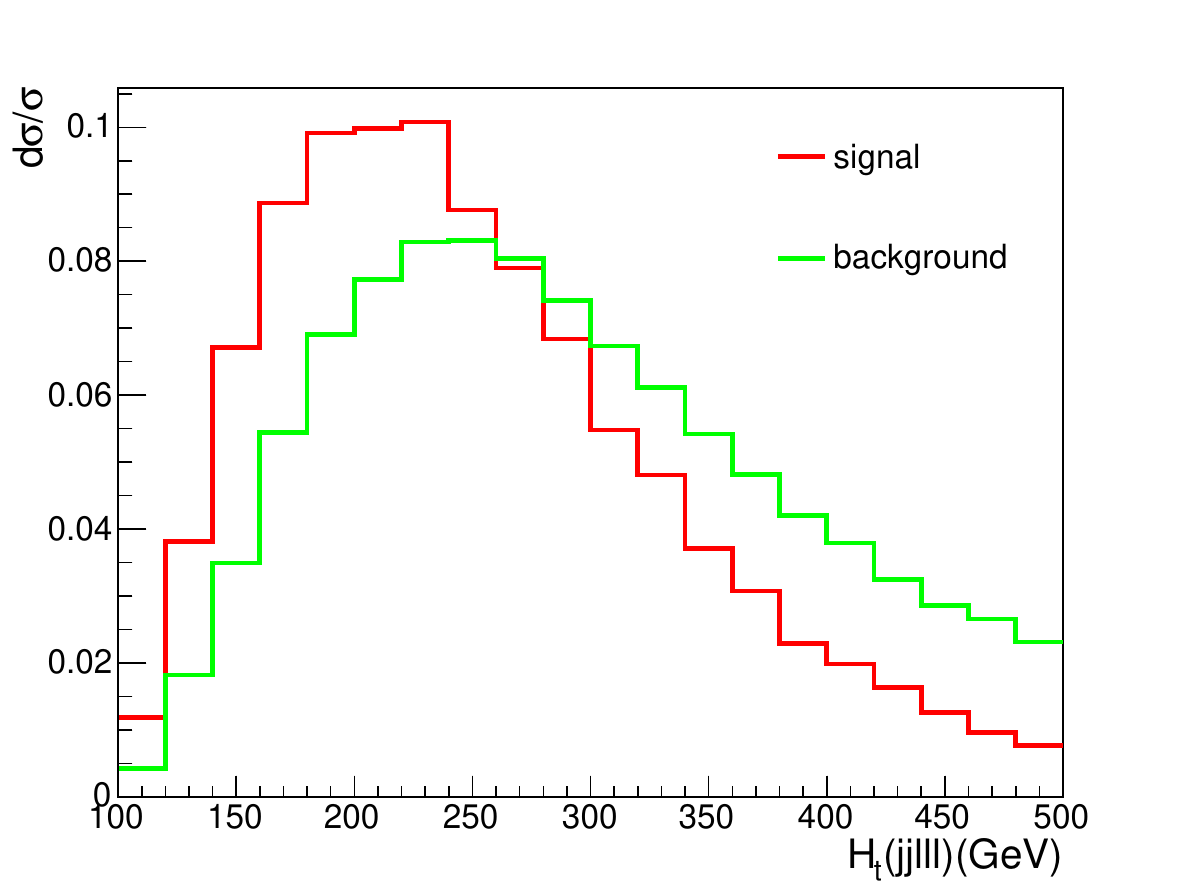}} 
\subfigure[~$P_T$ of harder jet in the hadronic W boson
\label{fig:j2pt}] 
{\includegraphics[width=.45\textwidth]{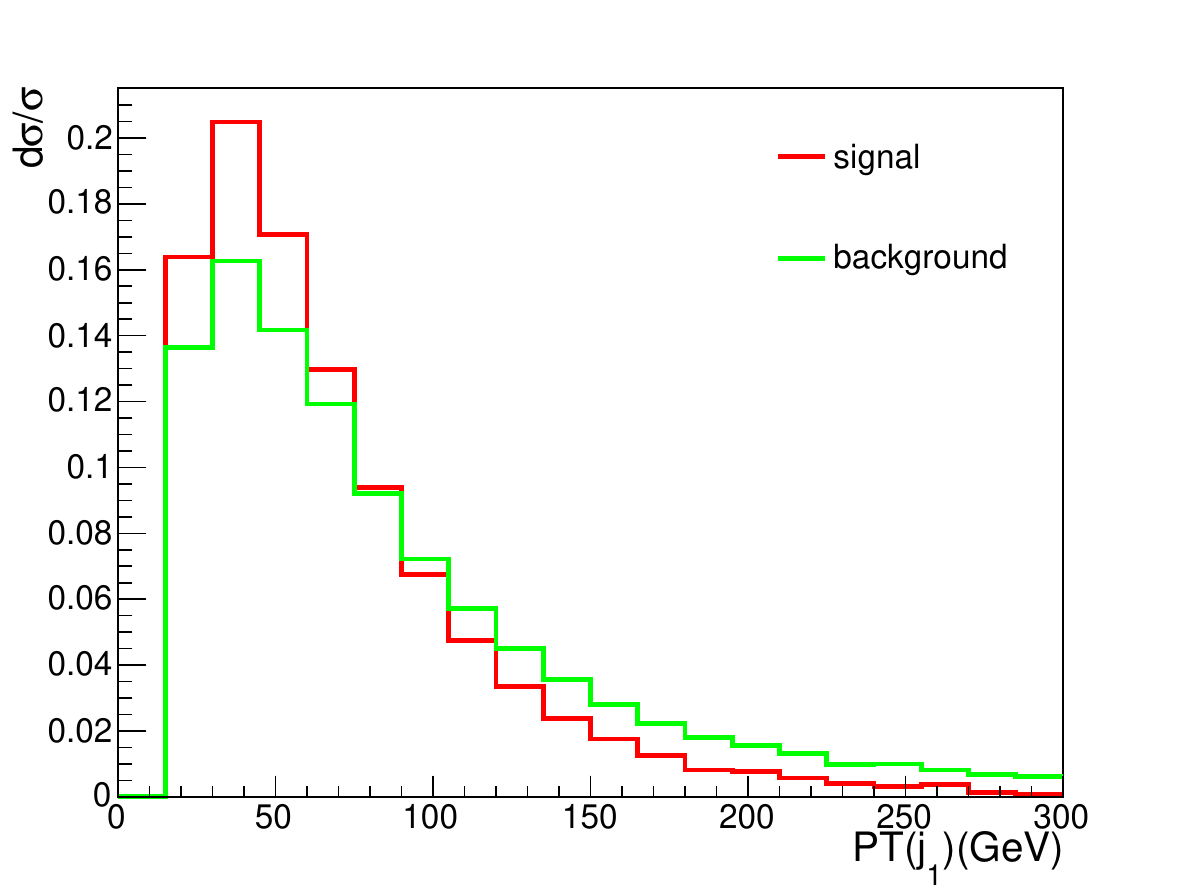}} \\
\subfigure[~Invariant mass of visible objects
\label{fig:l2pt}]
{\includegraphics[width=.45\textwidth]{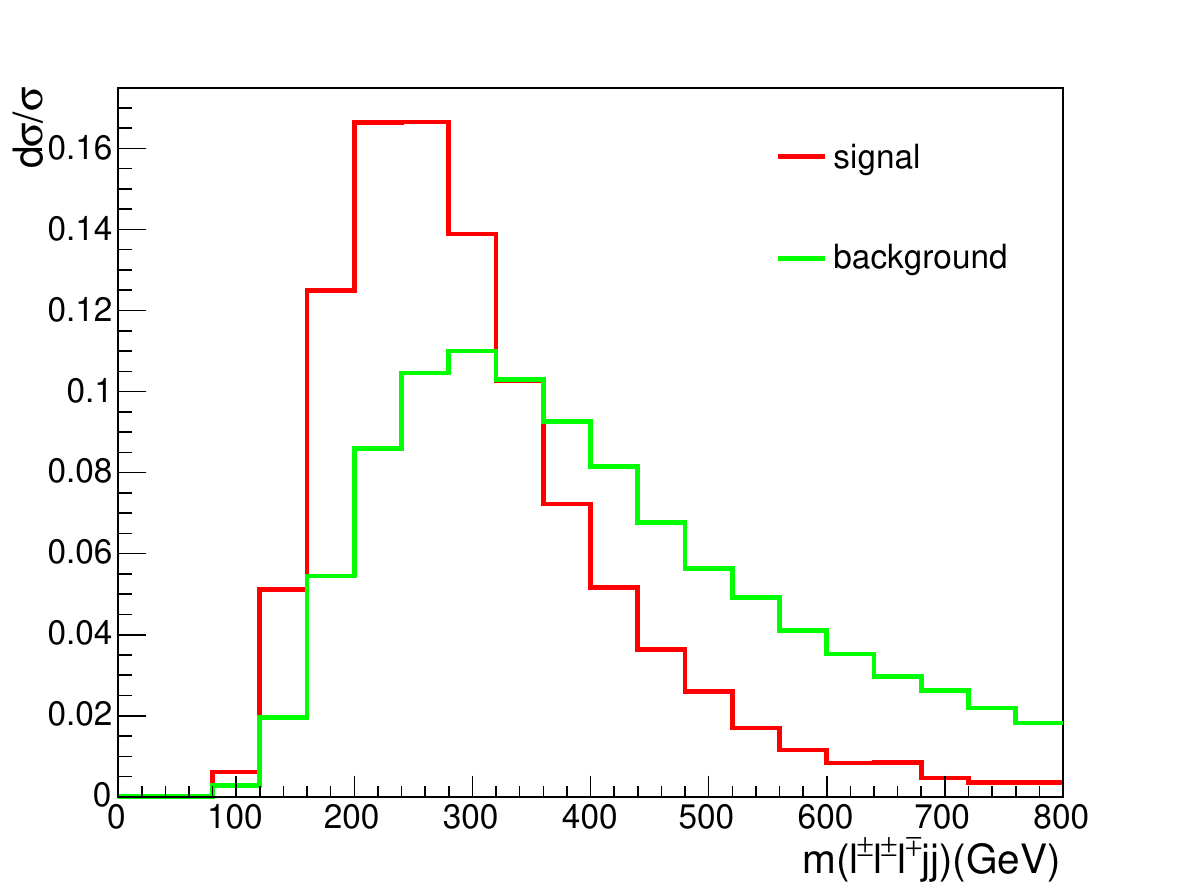}}
\subfigure[~The transverse mass of leptons and $\missE_T$
\label{fig:l3pt}] 
{\includegraphics[width=.45\textwidth]{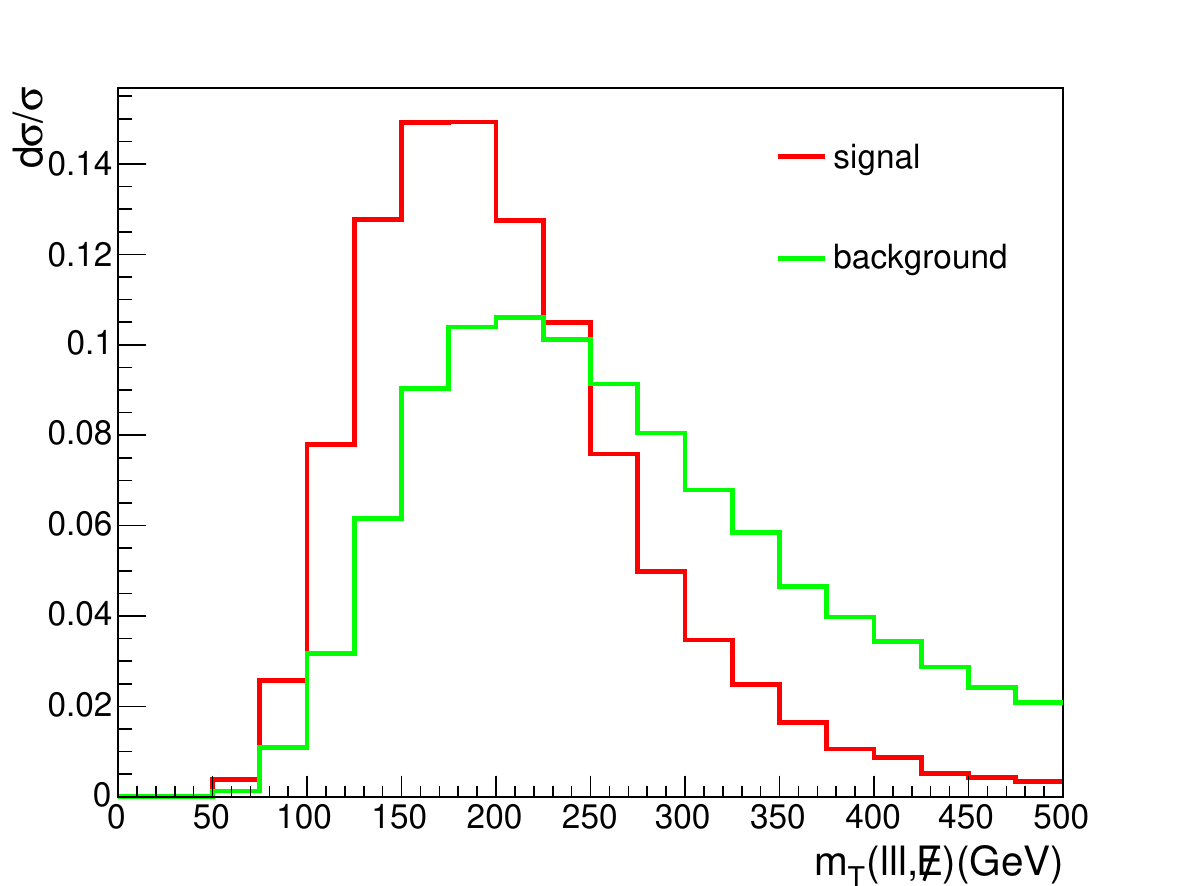}} 
\caption{We show more observables at hadron level used by the multivariable analysises.}
\label{fig:mva}
\end{figure*}

Considering that the number of signal event is few and the number of phase space of the final states is 24 (there are 9 dimensional space contributing to the missing energy in signal events) and most of variables are correlated, we optimize these cuts and include more observables which are independent of those four observables in the cut-based method. We have included more kinematic observables in our analysis:
\begin{itemize}
\item The sum of transverse momenta of all objects used in the reconstruction procedure is considered, of which the distribution of signal and background are shown in Fig. (\ref{fig:mva}a).
\item The transverse momenta of the harder jet used to reconstruct the hadronic W/W$^*$ is taken into account and is shown in Fig. (\ref{fig:mva}b). Due to the existence of off-shell W bosons, the momentum is softer than the background events.
\item The invariant mass of all the visible objects (including 3 leptons and all jets) is presented in Fig. (\ref{fig:mva}c), we observe that the signal events typically have smaller values when compared with the background events.
\item The transverse mass obtained from the combined 4-momentum of three leptons and the missing transverse momentum is shown in Fig. (\ref{fig:mva}d).
\end{itemize} 
We have also exploited other observables, like the transverse momenta of leptonic and semi-leptonic Higgs, the angular separation of two partially reconstructed Higgs bosons, the number of jets in each event, the ratio of missing energy over the visible energetic, etc.

We apply two multivariable analysis's: one is BDT, the other is MLP neural network. The distributions of other useful observables are shown in Fig. (\ref{fig:mva}). 

The results of multivariable analysis are presented at Table (\ref{table:TMVA1.0}) and the distributions of discriminant response to signal and background are shown in Figure (\ref{fig:nnbdt}). We observe that the $S/B$ can be improved by a factor $20$ and the significance can reach up to $2.0$ or so, which are very encouraging.

\begin{figure*}[!htbp]
\centering
\subfigure[~MLP Neutral Network
\label{fig:mlp}]
{\includegraphics[width=.45 \textwidth]{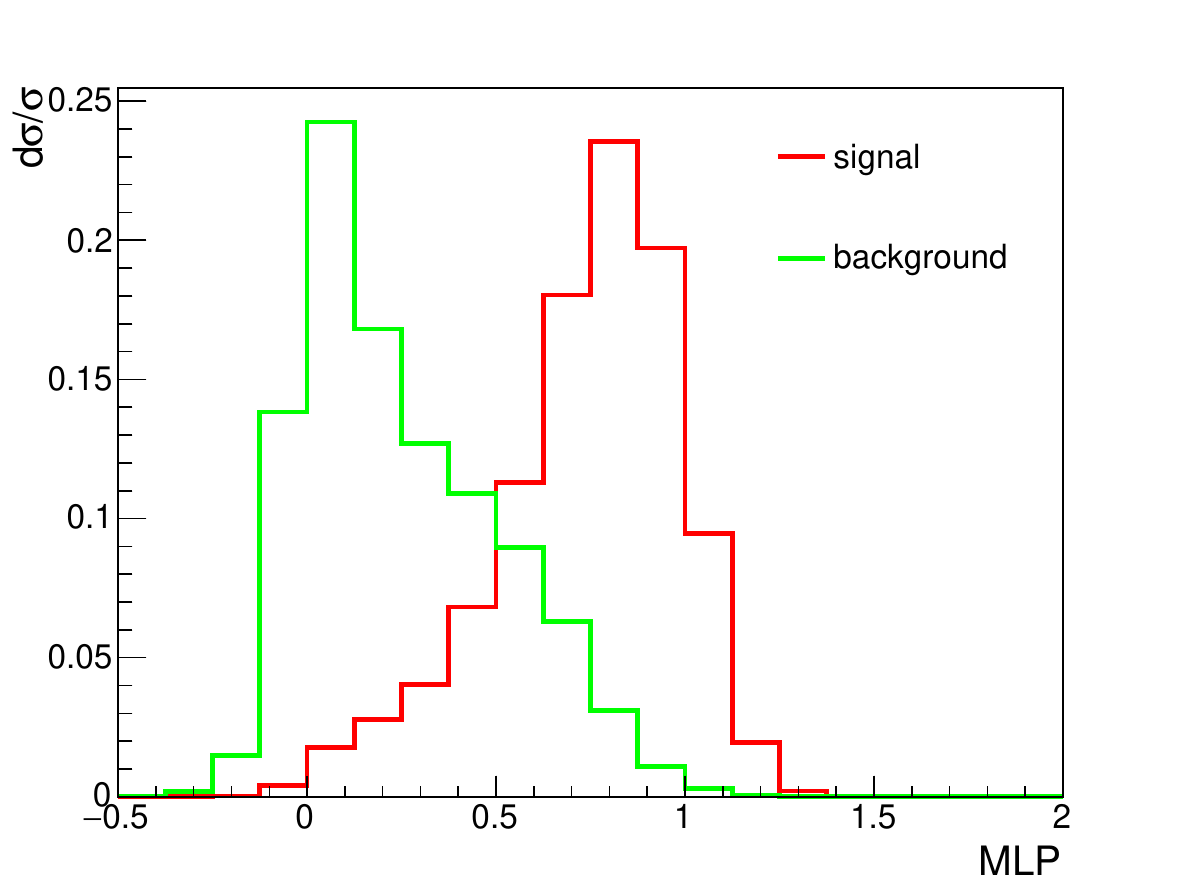}}
\subfigure[~Boosted Decision Tree
\label{fig:bdt}]
{\includegraphics[width=.45\textwidth]{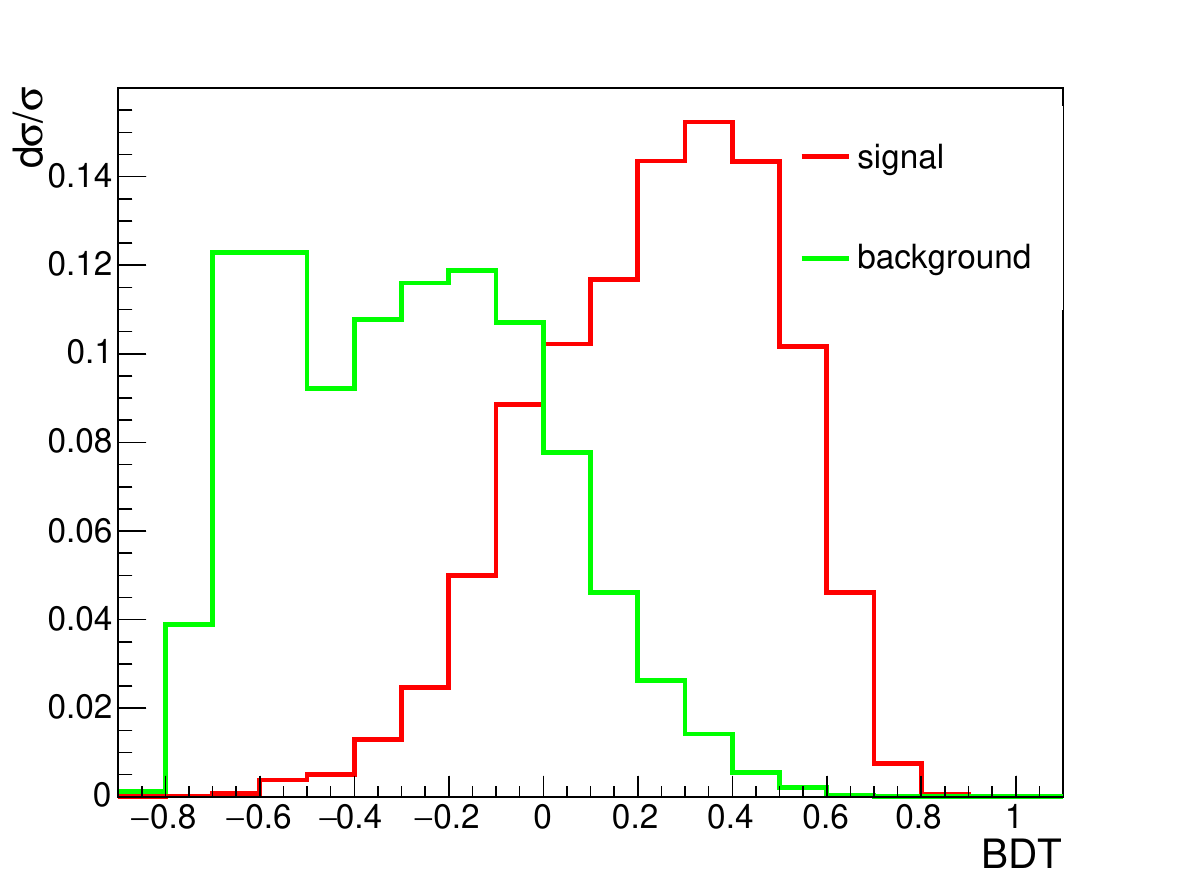}} 
\caption{The response of the discriminants to signal and background in two Multivariable Analysises, the MLP NN and BDT methods, are demonstrated.}
\label{fig:nnbdt}
\end{figure*}

\begin{figure*}[!htbp]
\centering
\subfigure[~Sensitive to $\lambda_3$ at 14 TeV LHC
\label{fig:s014}]
{\includegraphics[width=.45 \textwidth]{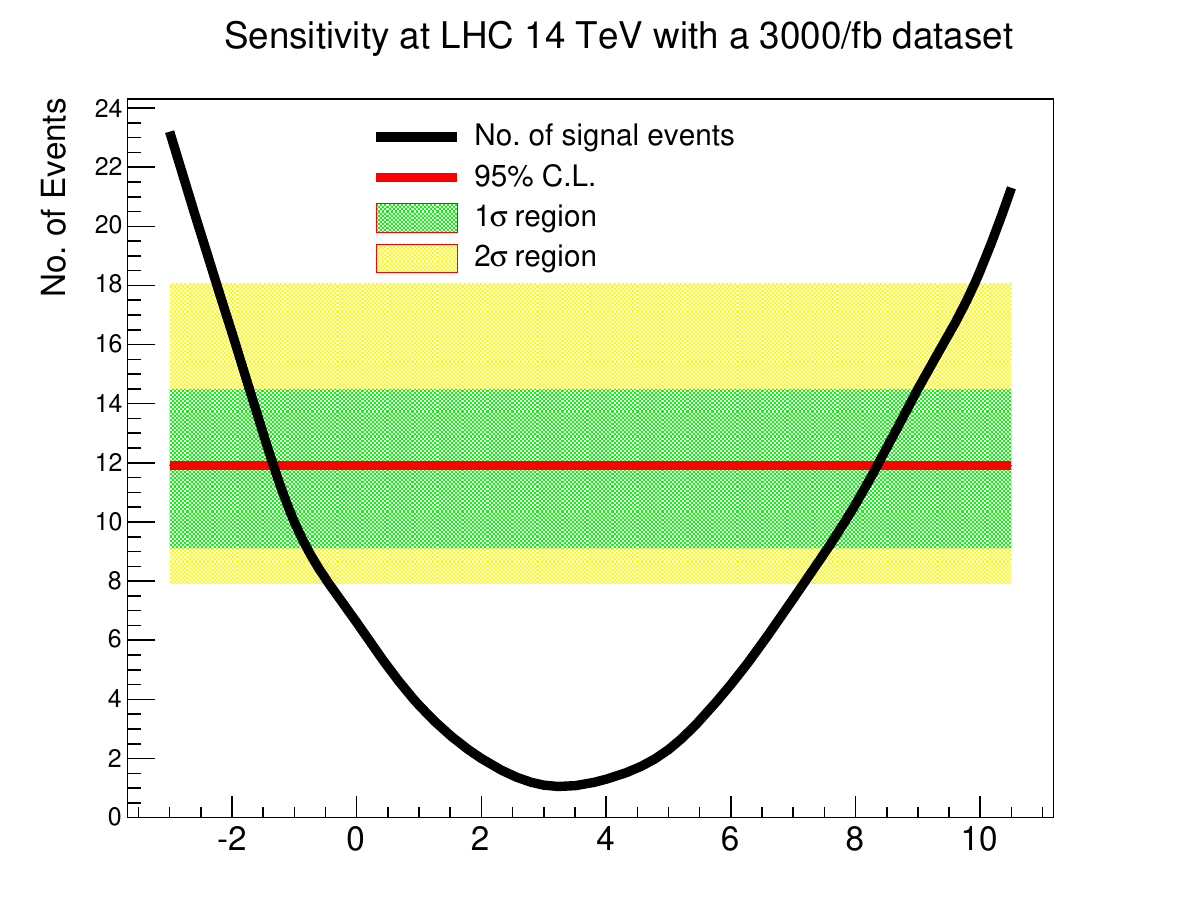}}
\subfigure[~Sensitivity to $\lambda_3$ at a 100 TeV collider
\label{fig:s100}]
{\includegraphics[width=.45\textwidth]{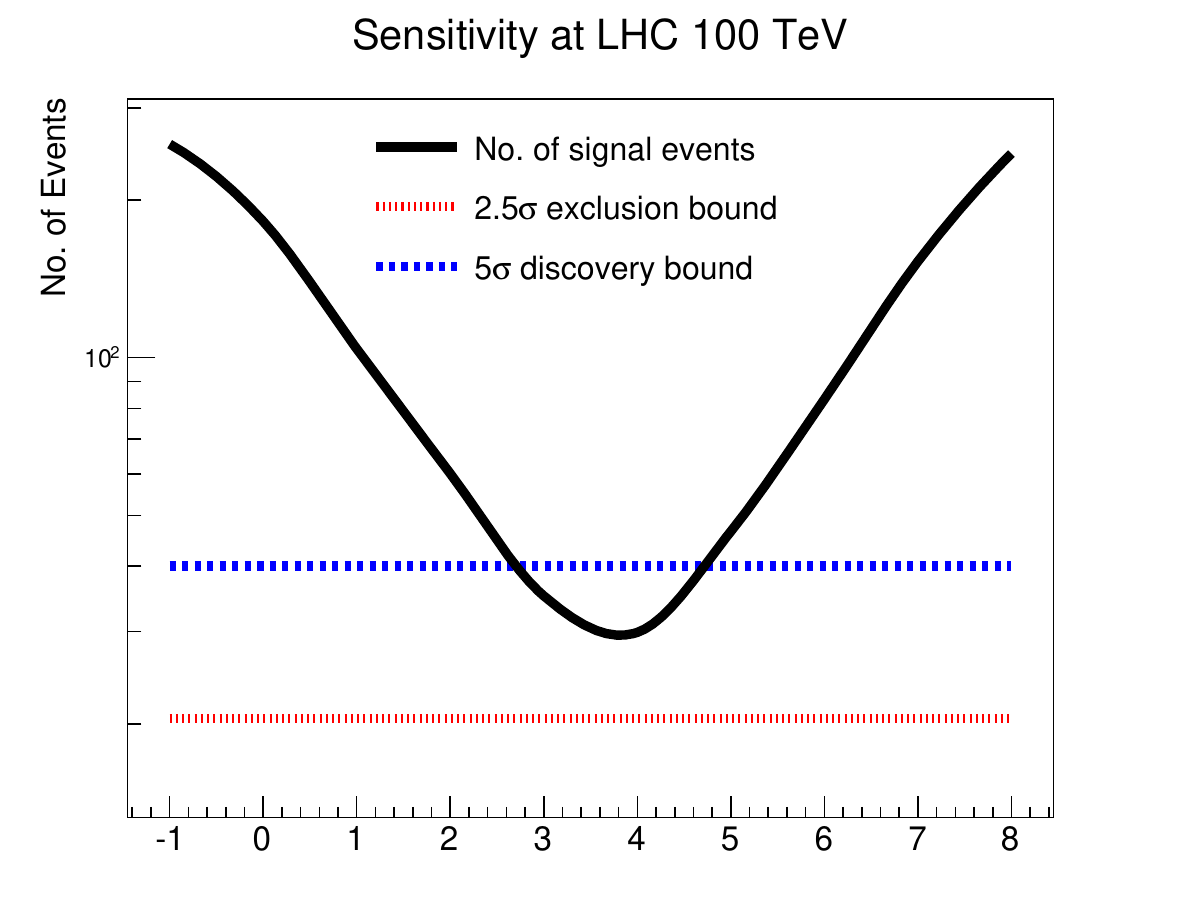}} 
\caption{The sensitivity to the triple couplings of Higgs boson, $\lambda_3$, at the LHC 14 TeV and a 100 TeV collider is shown.}
\label{fig:nnbdt}
\end{figure*}

We plot the estimated sensitivity to $\lambda_3$ at LHC 14 TeV with a 3 ab$^{-1}$ dataset in Fig. (\ref{fig:s014}). Although there are the large number of background events, we are capable to rule out the value of $\lambda_3<-1.0$ and $\lambda_3 > 8.0$; while if $\lambda_3$ is within the range $-1.0< \lambda_3<8$, it might be challenging to determine the value of $\lambda_3$ due to background fluctuations.

\subsection{The sensitivity to $\lambda_3$ at a 100 TeV Collider}
We apply our analysis demonstrated in the last section to a 100 TeV collider. It is noticed that both the production rate of signal and background with top quarks enhanced by a factor 40 or more than 100. In Table (\ref{table:cbm100TeV}), we tabulate the results obtained from the cut-based method and find that the significance can reach to $7.0$. 

When compared with the 14 TeV collider case, it is remarkable that the background $t \bar{t} H$ becomes the dominant one after all cuts. Here we haven't applied any special variable to further suppress this type of background, we expect better results when these special variables of $t \bar{t}H$ are used. Another remarkable fact is that although the production rate of the background from $t \bar{t} t \bar{t}$ also enhances by a factor 200, simply by counting the number of jets can efficiently kill most of this type of background.

\begin{table}[th]
\begin{center}%
\begin{tabular}
[c]{|c|c|c|c|c|c|c|}\hline
 & signal H H &  t $\bar{\textrm{t}}$ + W  & H W & W W W & t $\bar{t}$ H & $t \bar{t} t \bar{t}$ \\ \hline \hline
No. after preselection  &   $416.8$ &$4392.3$ & $716.1$ & $4384.1$ & $5045.9$ & $263.48$\\ \hline \hline
$m_{T2} < 110$ GeV &  $234.3$ &   $234.6$ & $116.8$ & $125.4$ & $1152.9$ & $26.8$\\ \hline \hline
$m_{H(\ell \ell)} < 55$ GeV & $202.3$ & $133.9$ & $94.8$ & $71.6$ & $811.6$ & $15.5$ \\ \hline \hline
No. of jets $<= 4$ & $160.0$ & $81.8$ & $82.4$ & $53.5$ & $304.9$ & $1.0$ \\ \hline \hline
$S/B$ &  \multicolumn{6}{c|} {0.31} \\ \hline
$S/\sqrt{B} $ & \multicolumn{6}{c|} {7.0}  \\ \hline
\end{tabular}
\end{center}
\caption{The effects of each cut in the cut-based method are demonstrated in a sequential way for a 100 TeV collider. After all cuts, the values of S/B and $S/\sqrt{B}$ are provided. }%
\label{table:cbm100TeV}%
\end{table}

In Table. (\ref{table:TMVA100TeV}), we tabulate the optimised results. Similarly to the 14TeV case, we notice that the significance can be improved by a factor of 3.5 or so and the ratio $S/B$ can be improved by two order.

\begin{table}[th]
\begin{center}%
\begin{tabular}
[c]{|c|c|c|c|c|}\hline
 & after preselection cuts &  Cut-based method & MLP method & BDT method  \\ 
 & & & $N_{NN} > 0.94 $ & $N_{BDT} > 0.22$\\ \hline
No. of Signal& $416.8$ & $160.0$ & $80.4$ & $104.0$ \\ \hline
No. of Background & $14801.8$ & $523.6$ & $107.3$ & $67.1$ \\ \hline
$S/B$ &  $2.8\times 10^{-2}$& $3.1 \times 10^{-1}$ & $7.5 \times 10^{-1}$ & $1.5$ \\ \hline
$S/\sqrt{B}$ &  $3.43$& $7.0$ & $7.8$ & $12.7$ \\ \hline
\end{tabular}
\end{center}
\caption{Comparison of significance among three analysis methods in a 100 TeV collider are shown.}
\label{table:TMVA100TeV}%
\end{table}

The sensitivity of a 100 TeV collider to the triple coupling $\lambda_3$ is provided in Fig. (\ref{fig:s100}). So a 100 TeV collider can exclude all values of $\lambda_3$ by simply using the 3 leptons mode considered in this work. Here our multivariate analysis has been optimised for the SM, i.e. $\lambda_3 =1$, all $\lambda_3$ out the range [2.8, 4.5] can be discovered. Nonetheless, if we optimise our analysis to different $\lambda_3$, we notice that even for the minimal cross section case with $\lambda_3 =3.6$ or so, the significance can reach to 5$\sigma$.

\begin{figure*}[!htbp]
\centering
\subfigure[~Three leptons mass after preselection
\label{fig:mlllpre}]
{\includegraphics[width=.45 \textwidth]{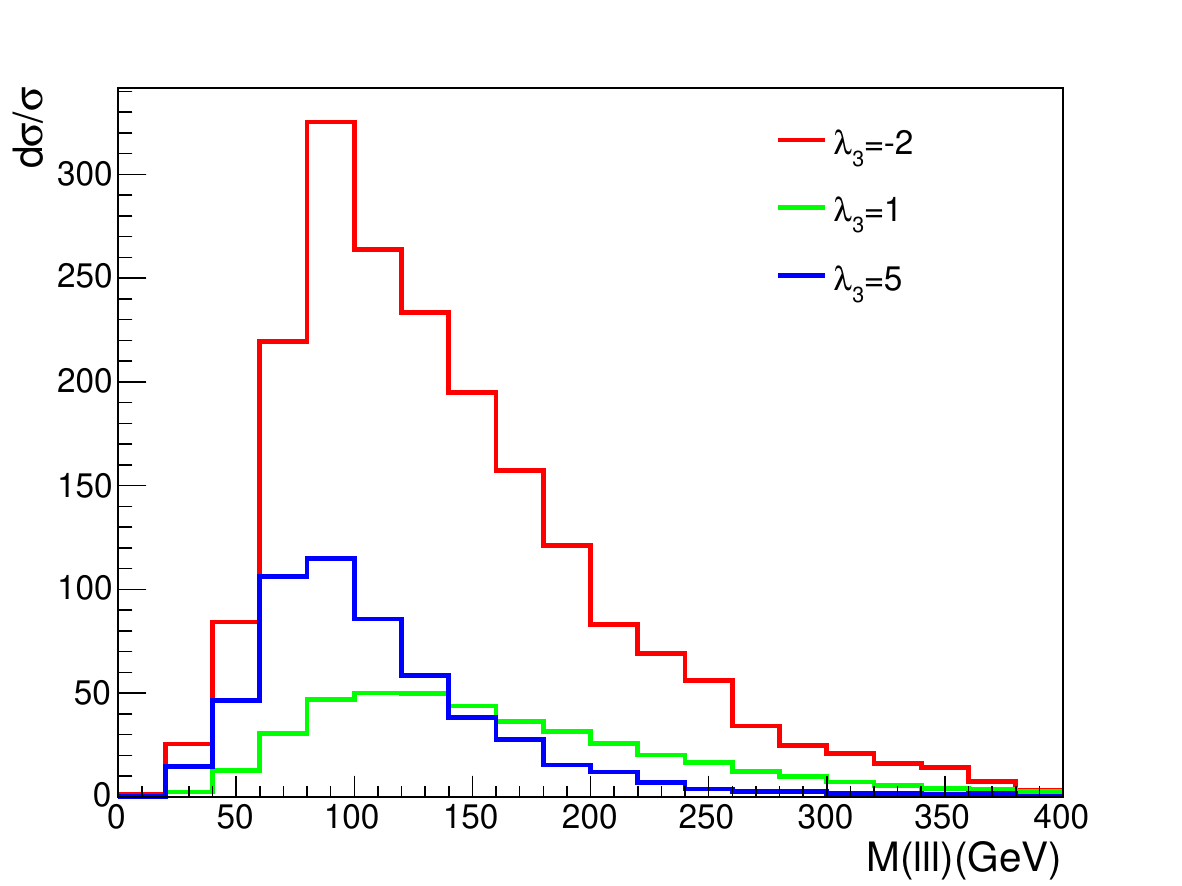}}
\subfigure[~Three lepton mass after MVA cut
\label{fig:mlllbdt}]
{\includegraphics[width=.45\textwidth]{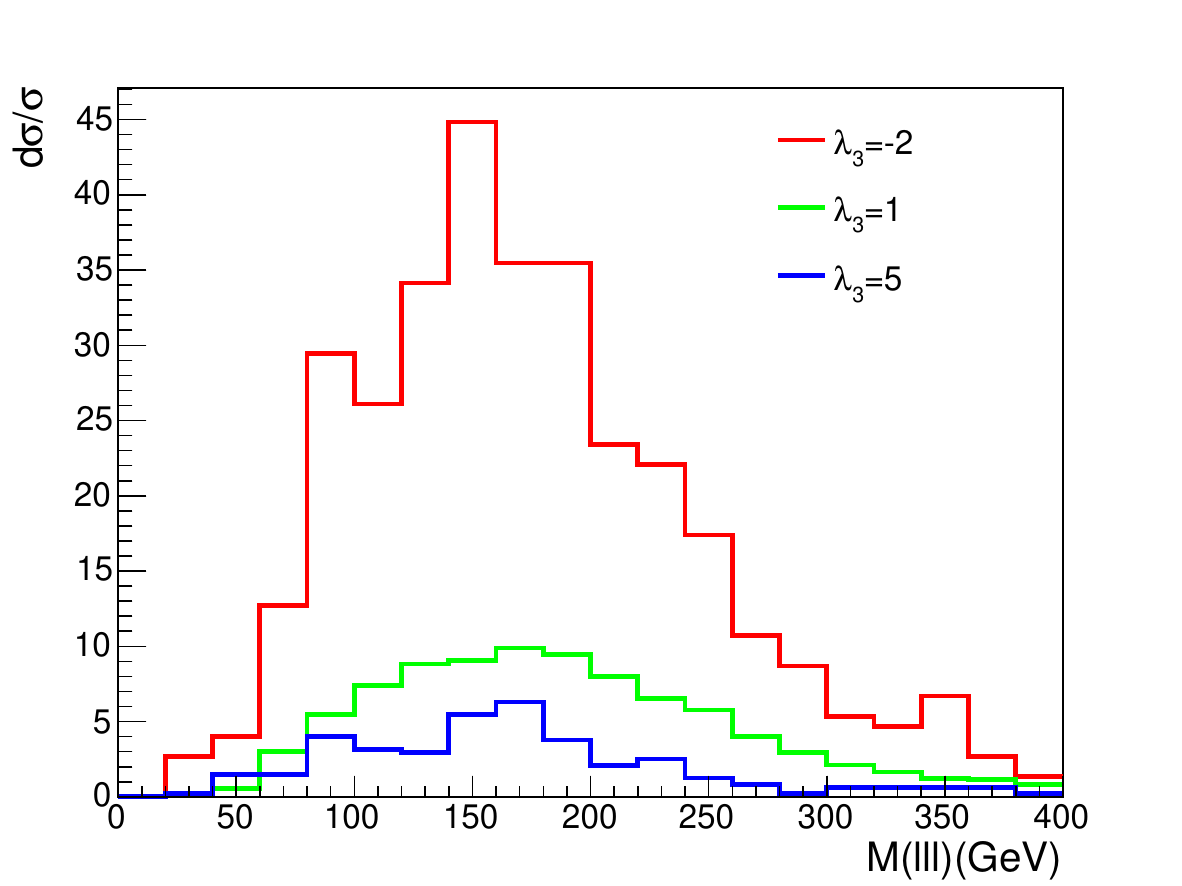}} 
\caption{The distributions of the invariant mass of three lepton after preselection and after MVA cut are demonstrated, where the background events have been neglected. Three different values of $\lambda_3$ are chosen to show how the triple Higgs coupling can affect the magnitudes and shapes.}
\label{fig:mlll}
\end{figure*}

Since there is no doubt that the SM triple Higgs coupling can be discovered at a 100 TeV collider, below we concern the issue how well this coupling can be measured by just using the trilepton mode considered in this work. To address this issue, we use the invariant mass of three leptons to perform a $\chi^2$ analysis. The distributions of this variable after the preselection cuts and the multivariate analysis are shown in Fig. (\ref{fig:mlll}). We have deliberately chosen three different values of $\lambda_3$ to demonstrate the differences in magnitudes and shapes. Since all cuts are optimised to the SM case, one can perceive the signal shapes of the cases $\lambda_3=-2$ and $\lambda_3=5$ have been greatly changed by the MVA filter.

\begin{figure*}[!htbp]
\centering
\subfigure[~$\chi^2$ varying with $\lambda_3$
\label{fig:chi2a}]
{\includegraphics[width=.45 \textwidth]{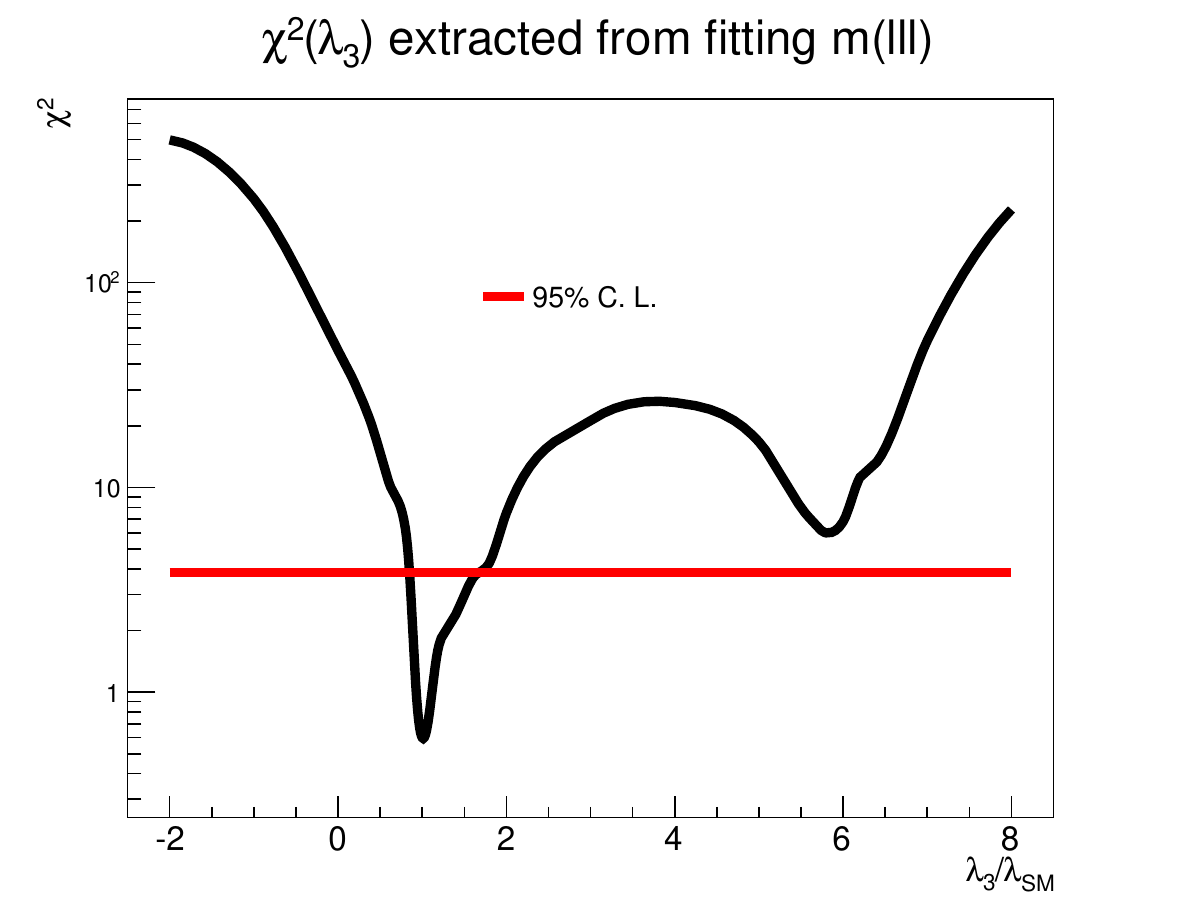}}
\subfigure[~Fine structure of $\chi^2$ near $\lambda_3=1$
\label{fig:chi2f}]
{\includegraphics[width=.45\textwidth]{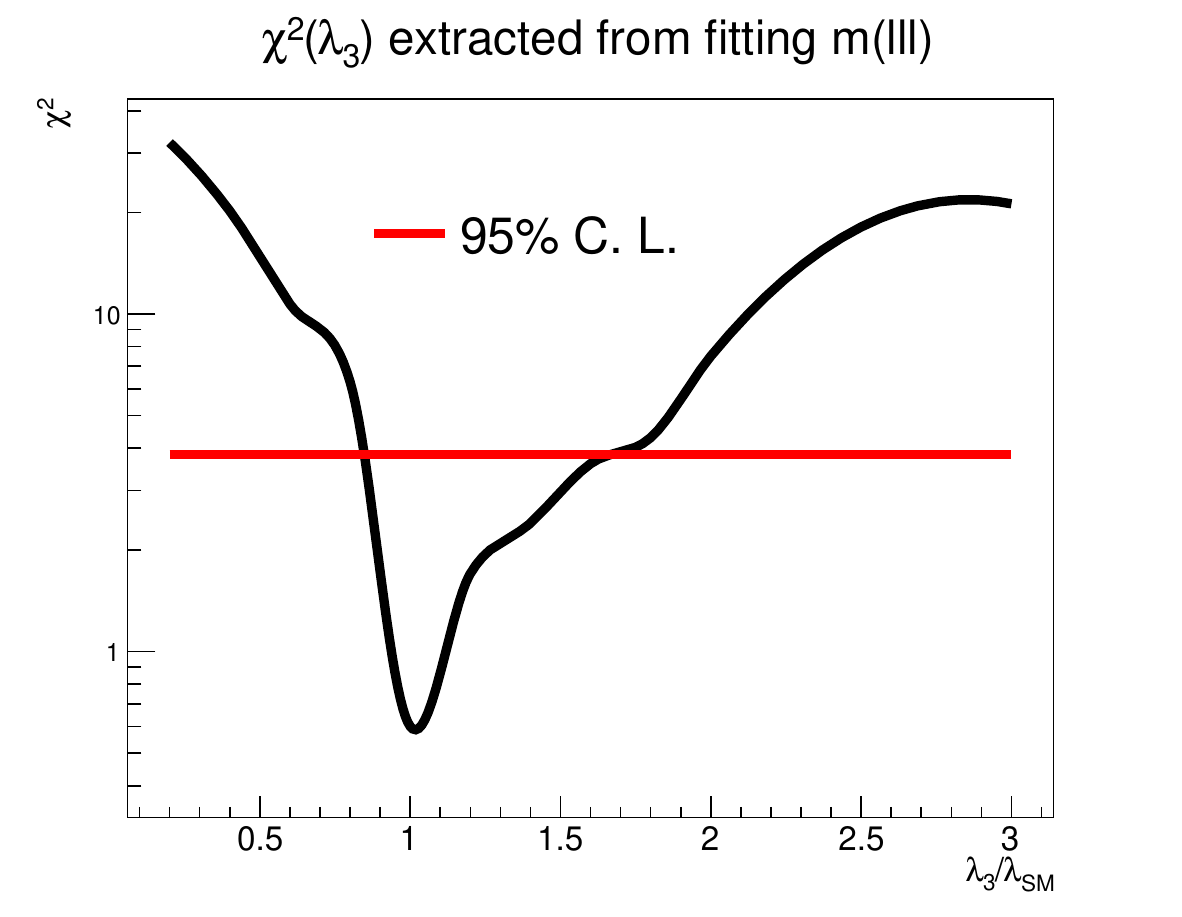}} 
\caption{The $\chi^2$ varying with $\lambda_3$ at a 100 TeV collider with a dataset 3000 fb$^{-1}$ is shown.}
\label{fig:chi2}
\end{figure*}

Below we address the issue how well the value of $\lambda_3$ can be determined. By using the method described in Refs. \cite{Baur:1992cd,Baur:2002qd}, we use 10 bins to perform a $\chi^2$ analysis on the distributions of the invariant mass of three leptons. The expression for $\chi^2$ is given by \cite{Baur:1992cd} 
\bea
\chi^2(\lambda_3) = \sum_{i=1}^{n_D} \frac{(N_i - f N_i^0 )^2}{f N_i^0} + (n_D-1)\,,
\eea
where $n_D$ denotes the number of bins, and $N_i$ is the number of events which include both the signal and background after cuts. Obviously $N_i$ is dependent upon the triple Higgs coupling parameter $\lambda_3$. $N_i^0$ means the number of events in the SM in the i-th bin after cuts, here the SM means $\lambda_3=1$. Here the quantity $f$ encodes the uncertainty in the normalisation of the SM cross section within the allowed range, which is determined by minimising $\chi^2$
\begin{displaymath}
f = \left \{  \begin{array}{ll} 
(1 + \Delta N)^{-1} & \textrm{for $\bar{f} < (1 + \Delta N)^{-1}$}, \\
\bar{f}  & \textrm{for $(1 + \Delta N)^{-1} < \bar{f} <1 + \Delta N$}\,, \\
1 + \Delta f  & \textrm{for $\bar f > 1 + \Delta N$}\,,
    \end{array} \right .
\end{displaymath}
where $\Delta N$ is taken as $10\%$ of the SM cross section (including both the signal and background after all cuts). The parameter $\bar{f}$ is defined as
\begin{displaymath}
\bar{f}^2 = \sum_{i=1}^{n_D} \frac{N_i^2}{N_i^0} / ( \sum_{i=1}^{n_D} N_i^0 )\,.
\end{displaymath}
The results are presented in Fig. (\ref{fig:chi2}) where the common term $n_D-1$ has been omitted. There are two comments in order:
\begin{itemize}
\item In the Fig. (\ref{fig:chi2a}), we observe that a 100 TeV collider can distinguish the two cases $\lambda_3=1$ and $\lambda_3=6.3$, although total cross sections of these two cases are equal. The difference in the lineshapes of the invariant mass of three leptons is sufficient to separate them from each other. 

To appreciate the underlying reason why these two cases are separable, at leading order, we can represent the differential cross section in the following form as 
\bea
\frac{d^2 \sigma}{d s\,\, d \cos\theta} &=& (\lambda_3 C_{\triangle} + C_{\Box} )^2 + D_{\Box}^2 \,,
\eea
where the $C_{\triangle}$ and $C_{\Box}$ are $S=0$ form factors and $D_{\Box}$ is $S=2$ form factor, which are dependent upon $s$ and $\cos\theta$ and their exact expressions at leading order can be found in \cite{Glover:1987nx}. In term of this form, total cross section can be expressed as 
\bea
\overline{ \sigma } = \lambda_3^2 \overline{C_{\triangle}^2} + 2 \lambda_3 \overline{C_{\triangle} C_{\Box}} + \overline{C_{\Box} ^2 + D_{\Box}^2} \,,
\eea
where all overlined quantities, like $\overline{C_{\triangle}^2}$, $\overline{C_{\triangle} C_{\Box}}$, and $\overline{C_{\Box} ^2 + D_{\Box}^2}$, denote the integrated values, which are just numbers. When total cross section is fixed as $\overline{ \sigma }$ by experimental measurements, there are two solutions of $\lambda_3$ which can yield the same total cross section. These two solutions can be expressed below as
\bea
\lambda_3^\pm &=& \frac{\overline{C_{\triangle} C_{\Box}}  \pm \sqrt{ (\overline{\sigma} - \overline{C_{\Box} ^2 - D_{\Box}^2}) \overline{C_{\triangle}^2} -  \overline{C_{\triangle} C_{\Box}}^2} }{  \overline{C_{\triangle}^2}}\,.
\eea
From these two solutions, the difference of differential cross section between them can be expressed as
\bea
\frac{d^2 \sigma^+}{d s\,\, d \cos\theta} - \frac{d^2 \sigma^-}{d s\,\, d\cos\theta} =2  (\lambda_3^- - \lambda_3^+) \,\, C_{\triangle} \,\,  (  C_{\Box} - \frac{\overline{C_{\triangle} C_{\Box}}}{\overline{C_{\triangle}^2}} C_{\triangle})\,.
\eea
It is noticed that two form factors, i.e. $C_{\triangle}$ and $C_\Box$, can  completely determined the difference of lineshapes of these two cases.
\item In the Fig. (\ref{fig:chi2f}), we observe that by using the $\chi^2$ fit of the invariant mass of three leptons, the value of $\lambda_3$ can be determined as $1^{+0.6}_{-0.3}$ in the $95\%$ confidence level, which is wider than the value $1^{+0.2}_{-0.1}$ or so when only the statistic accuracy is taken into account.  But if the total luminosity of can reach to 30 ab$^{-1}$, it is possible to reach this precision or better. This result is comparable to the precision which might be achieved from both the signals of $b\bar{b} \gamma \gamma$ final state and those of $b\bar{b} \gamma \gamma$ with a hard jet \cite{Barr:2014sga}.
\end{itemize}

Since observables constructed from leptons are expected to be more robust than those constructed from jets due to the contamination of underlying events and pileup of high luminosity run at future colliders, so we have used the invariant mass of three leptons to perform the $\chi^2$ fit. We can use the lineshape of the transverse momenta of leptonic Higgs, the invariant of leptonic Higgs, etc, to perform alternative $\chi^2$ analysis. It is noticed that they can yield the similar results.

\section{Discussion and Conclusion}
In this work, we have considered the feasibility of $3\ell 2 j + \missE$ mode to discover the signal of Higgs pair production at the LHC and a 100 TeV collider. We have proposed a partial reconstruction procedure to reconstruct two Higgs bosons in the final state and have examined the $m_{T2}$ observable in the hypothesis of pair production to discriminate the signal and background events. Although the production rate of signal events at the LHC 14 TeV 3000 fb$^{-1}$ is small, we have found that this mode can yield a significance around $1.5$ or so. For a 100 TeV collider with the same integrated luminosity, we noticed this mode can be used to determine the $\lambda_3$ and triple coupling of the Higgs boson of the standard model can be determined into the range $1^{+0.6}_{-0.3}$. If the total integrated luminosity is 30 ab$^{-1}$,  we estimated that it is possible to achieve $1^{+0.2}_{-0.1}$.

\begin{figure}[!htb]
\begin{center}
\includegraphics[width=0.9 \columnwidth]{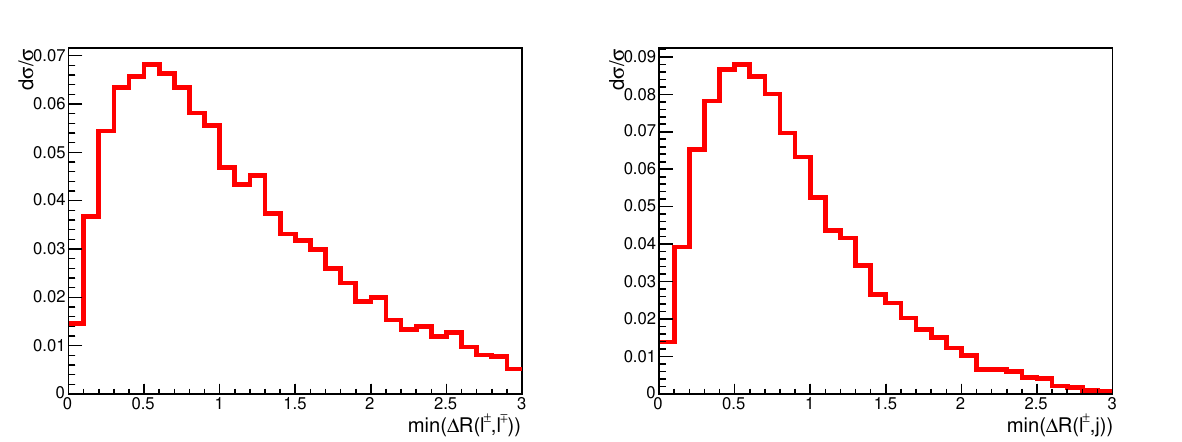}
\caption{For the signal events at the hadron level, we show the minimal angle separation between two leptons (the left panel) and the minimal angle separation between a lepton and a jet (the right panel) here.}
\label{partlllj}
\end{center}
\end{figure}

At our hadron level simulation, in order to take more signal into account, we have deliberately chosen $\Delta R(\ell) \geq 0.2$ to find isolated leptons; after having found these objects, we cluster the rest of particles and energy into jets. We notice that the angular separation cuts between leptons and that between a lepton and a jet can affect the selection efficiency of signal events to a quite considerable degree. Therefore, we provide the distributions of $\Delta R^{\textrm{min}} (\ell,\ell)$ and $\Delta R^{\textrm{min}} (\ell,j)$ in Fig. (\ref{partlllj}). 

In this work, we have used the lepton isolation criteria $\Delta R(\ell) =0. 2$, which is possible when the fine granularity of tracker detector and electromagnetic calorimeter are taken into account. By changing this condition to $\Delta R(\ell) = 0.3$, we observe that the signal loss is around $10\%$. When we demand $R^{\textrm{min}}(\ell, j) \leq 0.4$, we notice that the signal loss is around $15\%$. 

We haven't included more detailed detector effects, like the pileup effects which may mitigate the reconstruction of two soft jets coming from the off-shell W decay. For a more realistic 100 TeV collision study, the pileup effects can be a serious issue \cite{deFavereau:2013fsa} due to the fact that we need identify two jets in the signal event, which deserves our future careful study.

We have used the B veto and have assumed B tagging efficiency as $0.6$. If the B tagging efficiency is assumed to be $0.7$ and more background events from $t\bar{\textrm{t}} W$ can be better rejected, then we expect a better realistic significance. Beside, information of color-flow of two jets from W/$W^{*}$, which are colour singlet objects, can be used to determine the right combination and may provide further improvement. Considering these potential further improvements for this mode and, in contrast to the contamination of pileup effects and underlying events which might mitigate the modes with B jets in the final states for the $bb\gamma\gamma$ mode \cite{alt2014019}, we believe this mode might be robust and promising and should be seriously considered by experimenters. 

We can extend this work to study the same-sign dilepton modes of the Higgs pair production at both the context of the LHC and a 100 TeV collider. In a 100 TeV collider, the production rate of Higgs pair can be quite significant, we can extend the partial reconstruction method and analysis demonstrated here to the four-leptonic mode of Higgs pair production, which should be clean and robust against the contamination of underlying events and pileup effects.

\begin{acknowledgments} This work is supported by the Natural Science Foundation of China under the grant NO.  11175251, No. 11205008, No. 11305179, and NO. 11475180.
\end{acknowledgments}

\end{document}